\documentclass[journal]{IEEEtran}
\usepackage{graphicx}
\usepackage[justification=centering]{caption}
\usepackage{verbatim}
\usepackage{cite}
\usepackage{multirow}
\usepackage{colortbl,booktabs}

\usepackage{stfloats}
\usepackage{cite}
\usepackage{xcolor}
\ifCLASSINFOpdf
\else
\fi
\usepackage{amsmath}

\usepackage{amssymb}

\begin{document}
	
	\title{Morphological feature visualization of Alzheimer's disease via Multidirectional Perception GAN}

	\author{Wen~Yu,
		Baiying~Lei,
        Yanyan Shen,
        Shuqiang Wang,
		Yong~Liu,
		Zhiguang~Feng,
		Yong~Hu,
		Michael K. Ng
		\thanks{W. Yu, Y. Shen and SQ. Wang are with Shenzhen Institutes of Advanced Technology, Chinese Academy of Sciences, Shenzhen 518060, China. E-mail: sq.wang@siat.ac.cn; yy.shen@siat.ac.cn}
		\thanks{Baiying Lei is with School of Biomedical Engineering,  Health Science Center,Shenzhen University, Shenzhen 518055, China. E-mail:leiby@szu.edu.cn}
        \thanks{Yong Liu is with Gaoling School of Artificial Intelligence, Renmin University of China, Beijing, China}
		\thanks{Yong Hu is with Department of Orthopaedics and Traumatology, University of Hong Kong}
		\thanks{Zhiguang Feng is with College of Intelligent Systems Science and Engineering, Harbin Engineering University, 150001 Harbin, China}
		\thanks{Michael K. Ng is with Department of Mathematics, The University of Hong
			Kong, Pokfulam, Hong Kong.}
	}

	\markboth{IEEE TRANSACTIONS ON NEURAL NETWORKS AND LEARNING SYSTEMS }%
	{Shell \MakeLowercase{\textit{et al.}}: Bare Demo of IEEEtran.cls for IEEE Journals}
	
	\maketitle
	
	\begin{abstract}
		The diagnosis of early stages of Alzheimer's disease (AD) is essential for timely treatment to slow further deterioration. Visualizing the morphological features for the early stages of AD is of great clinical value. In this work, a novel Multidirectional Perception Generative Adversarial Network (MP-GAN) is proposed to visualize the morphological features indicating the severity of AD for patients of different stages. Specifically, by introducing a novel multidirectional mapping mechanism into the model, the proposed MP-GAN can capture the salient global features efficiently. Thus, by utilizing the class-discriminative map from the generator, the proposed model can clearly delineate the subtle lesions via MR image transformations between  the source domain and the pre-defined target domain. Besides, by integrating the adversarial loss, classification loss, cycle consistency loss and \emph{L}1 penalty,  a single generator in MP-GAN can learn the class-discriminative maps for multiple-classes. Extensive experimental results on Alzheimer's Disease Neuroimaging Initiative (ADNI) dataset demonstrate that MP-GAN achieves superior performance compared with the existing methods. The lesions visualized by MP-GAN are also consistent with what clinicians observe.
	\end{abstract}


	\IEEEpeerreviewmaketitle

	\section{Introduction}\label{introduction}

	\IEEEPARstart{A}{lzheimer's Disease (AD)}  is an irreversible and chronic neurodegenerative disease with progressive impairment of memory and other mental functions.   It is estimated to be the fifth leading cause of death in elderly people\cite{ADpriorRegion2}.  AD is caused by abnormal cell death in the brain, long before amnestic symptoms are observable\cite{ADfactor}.  The resulting brain atrophy is visible in structural magnetic resonance (MR) images.  To date, AD is incurable but preventable. Since machine learning tools have obtained great success in medical image computing \cite{wang2018bone, wang2020ensemble}, it is practicable to diagnose the early stages of AD by MR images for timely treatment\cite{ADnotCure2,  ADx1, ADx2} and many works \cite{lei2020deep, hu2019cross, hu2020brain, hu2020medical, yu2020multi, lei2022predicting, hu2021bidirectional} has been done to for the early stages of AD. Significant memory concern (SMC) and mild cognitive impairment (MCI)  are the transitional stages between normal controls (NC) and AD\cite{ADnotCure1}. SMC and MCI present mild symptoms, and the disease-related regions are very subtle in MR images.   Currently, the clinical diagnosis procedure is time-consuming and requires extensive clinical training and experience for clinicians. Thus, developing automatic methods by utilizing deep learning to visualize the brain changes for the early stages of AD  is highly desirable.  It can assist clinicians for  AD analysis and may provide meaningful information on the pathogenesis of cognitive decline.  However, this is a  challenging task due to several reasons, such as the low-intensity contrast between the lesion and other neighboring regions, the indistinct boundary of the lesion, and the irregular lesion shape.

	To visualize features of different Alzheimer's stages in MR images, there already exist several feature visualization methods based on classification.  These methods can be classified into two categories. (1) The Regions Of Interest (ROI)-based classification approaches\cite{ADpriorRegion2, ADregressionVisual, ADclassificationVisual, ADregressionVisual99} and patch-based classification  approaches\cite{ADvisual1}.  The performance of these methods is limited since the brain ROIs or patches need to be selected based on anatomical brain atlases or biological prior knowledge beforehand. Multiple steps are required to exact features from ROIs or patches for classification and subsequent visualization. Therefore, they tend to be sensitive to parameters and time-consuming;  (2) Three strategies to visualize features for a convolutional neural network (CNN) classifier. (i)By editing an input image and observing its effect on the prediction results, the occluded regions which have a significant impact on prediction can be visualized;  (ii)By analyzing the gradients of the prediction for an input image, a heatmap can be produced for visualization; (iii)By analyzing the activations of the feature maps for the image, the regions which are responsible for making the specific prediction can be visualized.   These classification-based feature visualization methods make their predictions based on local regions most relevant to the particular prediction rather than the whole image, and it may ignore features with low discriminative power if stronger features for the prediction are available.   As a result, if there is evidence for a category at multiple locations in the image (such as multiple AD lesions in MR images),  some lesions with low discriminative power may be ignored.  Moreover, visual features strongly depend on the performance of the classifier, and a large number of labeled samples are required to train a robust model.

	\begin{figure*}
		\centering
		\includegraphics[width=\linewidth]{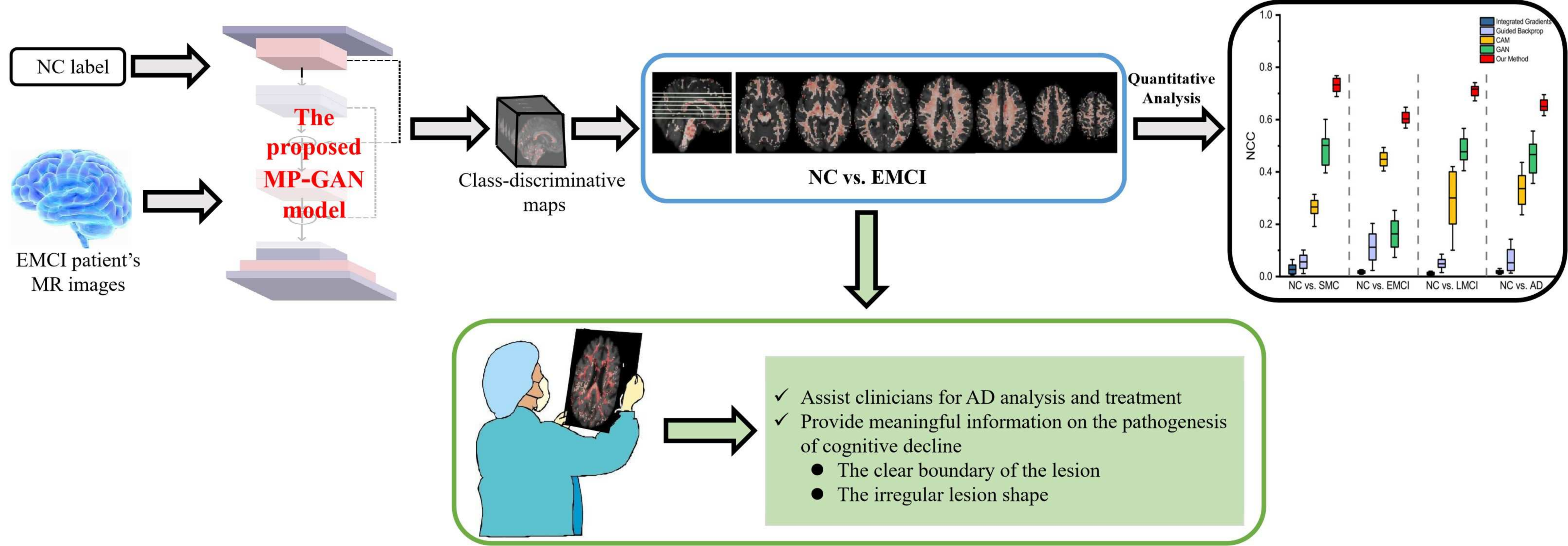}
		\caption {The pipeline of the MP-GAN model. Assuming the MP-GAN is well trained, given the real EMCI patient’s MR images and the NC label, the subtle morphological features between NC and EMCI can be visualized by the model to assist clinicians for AD analysis and treatment.}
		\label{fig_add_figure}
	\end{figure*}
	
	To alleviate these issues,  a novel  Multidirectional Perception Generative Adversarial Network (MP-GAN) is proposed to visualize morphological features for whole-brain MR images as shown in  Fig. \ref{fig_add_figure}. Generative Adversarial Network (GAN)\cite{gan_goodfellow, GanReview},deriving from variational methods \cite{mo2009, wang2009variational, wang2008variational}, has attracted lots of attention as it is capable of generating realistic data without explicitly modeling the probability density function. Specifically, the generator of   MP-GAN takes inputs as both MR images and its target domain.  Then it flexibly learns a class-discriminative map for the target domain.  By adding the class-discriminative map and the input MR image of the source domain, a synthetic MR image of the target domain can be produced. Thus the learned class-discriminative map can capture all brain changes by transforming the MR image between the source domain and the target domain.  By visualizing the class-discriminative maps,  the subtle and complex lesions that may not be found within one region can be identified.  Besides, by designing the hybrid loss, a single generator in MP-GAN can learn the class-discriminative maps for multiple-classes. In this manner, the common features unrelated to the specific domain can be reused during training, therefore the visualization performance is further improved.    With this global lesion visualization, clinicians can better exclude undesirable biases and potentially even identify previously unknown characteristics of AD.  To the best of our knowledge, the proposed MP-GAN is the first work to visualize the morphological features for different Alzheimer's stages by a single generator.   The contributions of this paper are summarized as follows:
	\begin{enumerate}
		\item      A novel MP-GAN with a multidirectional mapping mechanism is proposed to capture the salient global features efficiently. By utilizing the class-discriminative map from the generator, the proposed model can clearly delineate the subtle lesions via MR image transformations between the source domain and the target domain.
		\item  By integrating the adversarial loss, classification loss, cycle consistency loss and \emph{L}1 penalty, a single generator in MP-GAN can learn the class-discriminative maps for multiple-classes. The morphological features indicating different Alzheimer's stages can be visualized by a single MP-GAN model.
	\end{enumerate}
	
	The rest of this paper is organized as follows. The related work is reviewed in Section \ref{sec_Related_work}.  The proposed MP-GAN is described in detail in Section \ref{sec_methods}. In Section \ref{sec_Experiments},  MP-GAN is tested and compared with existing feature visualization methods to demonstrate its advantage. Finally, concluding remarks and future work are discussed in Section \ref{sec_Disscussion} and Section \ref{sec_Conclusion}.

	\section{Related work} \label{sec_Related_work}
	\textbf{Generative Adversarial Networks.}  GAN has attracted lots of attention as it is capable of generating realistic data without explicitly modeling the probability density function. It has shown remarkable results in various computer vision tasks such as image generation\cite{gan_goodfellow}, image-to-image translation  \cite{cycle_consistent_loss2}, image super-resolution \cite{GanReview}, and semi-supervised learning \cite{myGAN} \cite{GANgan}. A typical GAN model consists of two modules: a discriminator and a generator. The discriminator learns to distinguish between real and fake samples, while the generator learns to generate fake samples that are indistinguishable from real samples. Training the original GAN, however, suffers from several problems such as low quality of generated images, convergence problems, and mode collapse. To address these deficiencies, variants of the GAN were introduced\cite{wGAN,wGAN2}. The most representative work is Wasserstein GAN (WGAN) \cite{wGAN}. It leverages the Wasserstein distance to measure the distance between two data distribution that has better theoretical properties than the original KL (Kullback–Leibler) divergence.
		
		\textbf{Feature visualization methods.}The current feature visualization methods for AD generally fall into two categories:  (1)The ROI-based classification approaches; (2)The CNN-based classification approaches.
		
		For the first category, the brain ROIs or patches were selected based on anatomical brain atlases or biological prior knowledge beforehand, then multiple steps were required to extract features from ROIs or patches for classification. According to classification performance,  the most frequently selected ROIs or patches would be visualized\cite{ADregressionVisual, ADclassificationVisual}. For instance,  Lian et al. \cite{ADvisual1}  proposed a hierarchical fully convolutional network (H-FCN) to automatically identify discriminative local patches and regions in MR images for AD analysis.  The hierarchical discriminative locations of brain atrophy at both the patch-level and region-level were visualized.

		For the second category,  there were three strategies to visualize features for CNN.
		
		\begin{enumerate}
			\item      By editing an input image and observing its effect on the prediction results, the occluded regions which had a significant impact on prediction can be visualized\cite{swapTest}.  For instance, Zeiler and Fergus  \cite{visualCNN} proposed an occlusion-based method to visualize the activity within CNN. Different portions of the input image were occluded with a grey square, and the output of the classifier was observed.   The occluded regions which cause the probability of the correct class drop significantly would be visualized.  Korolev et al. \cite{classificationVisualResults}  utilized 3D-ResNet for AD classification, and the important regions of the MR image most affected by AD were visualized by the occlusion-based method \cite{visualCNN}.
			\item  By analyzing the gradients of the prediction for an input image, the heatmap can be produced for visualization \cite{gradient1,gradient2,gradient69,visualClassifierGradCAMICCV,gradientAttribution,ADvisual7}. For example, Springenberg et al. \cite{GuidedBackprop}  proposed a new variant of the ``deconvolution approach''  guided backpropagation for visualizing features learned by CNNs. Guided backpropagation can be applied to a broader range of network structures.  Sundararajan et al. \cite{IntegratedGradients}  proposed  Integrated Gradients by utilizing an axiomatic framework for feature visualization;
			\item  By analyzing the activations of the feature maps for the image, the regions which were responsible for making the specific prediction can be visualized. For instance,  Zhou et al.\cite{CAM} proposed Class Activation Mapping (CAM)  to visualize the discriminative object parts detected by CNN in a single forward pass.    Khan et al. \cite{ADCAM} utilized VGG with transfer learning for AD  analysis.  CAM was utilized to visualize the discriminative regions in the MR image for model interpretation. Lian et al. \cite{ADvisual2} proposed a multi-task weakly-supervised attention network (MWAN) by leveraging a fully-trainable dementia attention block for regression. The attention maps were visualized by CAM for AD subjects.  Sarraf et al.\cite{ADfeature} utilized LeNet-5 to classify structural MR images for AD vs. NC. The filters and the features were visualized for interpretation.
	\end{enumerate}
	
	\section{The Proposed MP-GAN}\label{sec_methods}
	
	\subsection{Overview}
\begin{figure*}
		\centering
		\includegraphics[width=0.9\linewidth]{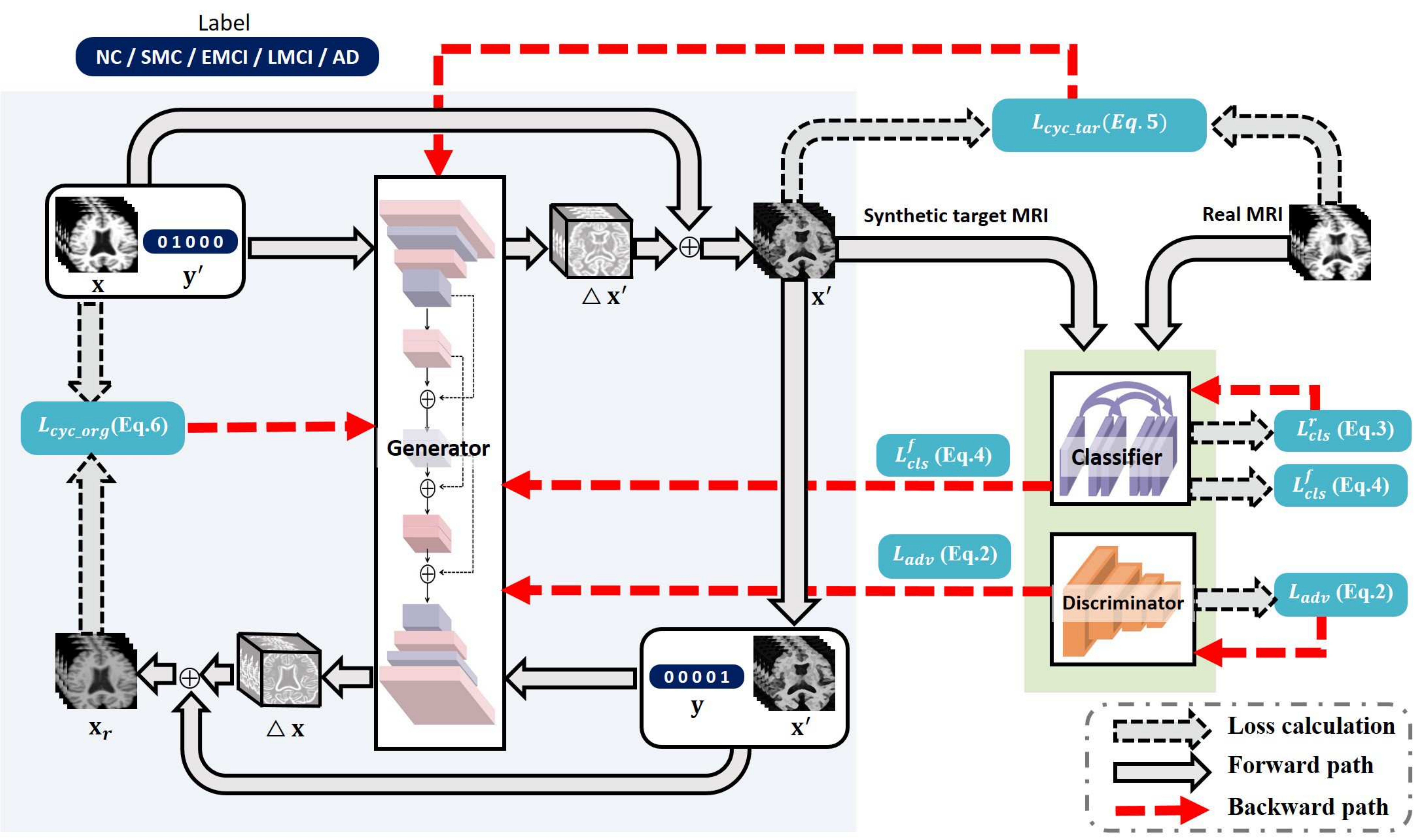}
		\caption{The flowchart of MP-GAN. It consists of three components: a generator, a classifier, and a discriminator. The generator maps the input source MR image to the synthetic target MR images by the class-discriminative map. The synthetic source MR images are reconstructed from the synthetic target MR images by the generator in the same manner. By utilizing classification loss, adversarial loss, and cycle consistency loss, the generator learns to generate synthetic target MR images and the reconstructed source MR images that are indistinguishable from real MR images.}
		\label{fig_FV-GAN}
	\end{figure*}

	This paper proposes a novel mapping mechanism by which the MR images can be mapped between each pair of source class and target class in a multidirectional manner. Take the source class NC as an example, the generator can map the MR images between NC and SMC, meanwhile, it can also map MR images between NC and the other class such as EMCI, LMCI, and AD simultaneously. The flowchart of MP-GAN is shown in Fig. \ref{fig_FV-GAN}. After data preprocessing (see Section \ref{sec_Datasets}),  the normalized T1-MR images of all classes are fed into MP-GAN. The proposed model learns the class-discriminative maps between all class-pairs for visualizing morphological features. More specifically, the generator aims to capture salient global features in class-discriminative maps. Then the class-discriminative maps are used to transform MR images between the source domain and the target domain.  To control semantic information, an auxiliary classifier is introduced based on the generator and discriminator to form the MP-GAN architecture. While the generator produces the class-discriminative maps distinguishing between the source domain and the target domain, the classifier predicts the domain indicating Alzheimer's stage, and the discriminator identifies whether the transformed MR images are real or fake. In this manner, the class-discriminative maps learned by MP-GAN can highlight exactly which regions of the MR image are significant for discrimination between the source domain and the target domain at the voxel-level.
	The subtle and complex lesions that may not be found within one region can be identified.  Furthermore, since the input MR images are high-order with complicated brain structure, MP-GAN is further designed with the following two improvements: (1) 3D Residual Blocks are exploited in the conditional generator so that the features from the low-level can be reused, and the vanishing-gradient problem can be prevented; (2) 3D-DenseNet is utilized in classifier to capture more discriminative features.
	
	\subsection{The Architecture}\label{sec-FVT-GAN}
	
	The proposed MP-GAN is designed to visualize morphological features for multiple-classes. To achieve this, the generator G is designed to produce class-discriminative map $\Delta x$ which can transform an input MR image $x$ to an output MR image  $x^{\prime}$ conditioned on the target class $y^{\prime}$, $[G(x, y^{\prime})+x] \rightarrow x^{\prime}$. During training, the target class   $y^{\prime}$ is randomly selected so that G learns to produce class-discriminative maps for all class-pairs.    By doing so,  the target class    $y^{\prime}$ can be predefined, and global features that distinguish between the source domain  $y$ and the desired target domain $y^{\prime}$ can be visualized at the testing stage.

	As illustrated in Fig. \ref{fig_FV-GAN}, input MR image $x$ is labeled and $y$ represents the corresponding class. The conditional generator aims to capture all salient global features in class-discriminative maps $\Delta x$.  Then $\Delta x$ is utilized to transform input MR image from the source domain $y$ to the target domain $y^{\prime}$ in a bidirectional manner.  The classifier predicts label $y_{c}$ given real MR image $x$ by the conditional distribution $p_{c}(y | x)$, and the discriminator is trained to identify whether the MR image is real or fake.   Formally, given an MR image  $x$ of source class   $y$ and a conditional variable $y^{\prime}$, the generator can produce a synthetic MR image $x^{\prime}$ of target class $y^{\prime}$ by adding the generated class-discriminative map $\Delta x$ and input MR image $ x$.
	\begin{equation}
		x^{\prime}=\Delta x+x =G\left(x,  y^{\prime}\right)+x,
	\end{equation} which is indistinguishable from the real MR image of the target domain $y^{\prime}$. Thereby, class-discriminative map  $\Delta x$ contains all salient global features which distinguish between two domains $y$ and $y^{\prime}$.  The change of salient voxels between the source domain $y$ and the target domain $y^{\prime}$  on the MR image can be visualized by the class-discriminative map.
	\\
	
	\textbf{Adversarial Loss. }     To make the synthetic target MR images indistinguishable from real MR images,  an adversarial loss is defined as
	
	\begin{equation}\begin{aligned}
			\mathcal{L}_{a d v}=& \mathbb{E}_{x}\left[\log D(x)\right]+\\
			& \mathbb{E}_{x, y^{\prime}}\left[\log \left(1-D(G\left(x,  y^{\prime}\right)+x)\right)\right],
		\end{aligned}
		\label{equ_adv}
	\end{equation} where generator G generates an MR image $[G\left(x,  y^{\prime}\right)+x]$ conditioned on both the input MR image $x$ and the target class $y^{\prime}$, while discriminator D attempts to distinguish between real and fake MR  images.  The G tries to minimize this adversarial loss, while the D tries to maximize it. More specifically, when the discriminator successfully identifies real and fake MR images, it is rewarded and no change is needed to update the parameters of the discriminator, whereas the generator is penalized with large updates to parameters. Alternately, when the generator fools the discriminator, it is rewarded, and no change is needed to update the parameters of the generator, but the discriminator is penalized and its component parameters are updated.
	\\
	
	\textbf{Classification Loss.}  Given an input MR image  $x$ and a target class  $y^{\prime}$,  the goal of  MP-GAN  is to produce a class-discriminative map that can transform $x$ into an output MR image $x^{\prime}$.  $x^{\prime}$ aims to be classified as the target class $y^{\prime}$. To achieve this condition,  an independent classifier is introduced and  the classification loss is imposed when optimizing generator G. Specifically, the  loss function is decomposed into two terms: a  classification loss of real images to optimize classifier C, and a classification loss of fake images to optimize  generator G. In detail, the former is defined as
	\begin{equation}
		\mathcal{L}_{c l s}^{r} =\mathbb{E}_{\left(x, y\right) \sim p_{\text {real }}(x, y)}\left[-\log p_{c}\left(y | x\right)\right].
	\end{equation}
	By minimizing this classification loss,  classifier C learns to classify a real  MR image $x$ to its corresponding class $y$.  On the other hand, the loss function for the  classification of fake images is defined as
	\begin{equation}
		\mathcal{L}_{c l s}^{f}=\mathbb{E}_{\left(x^{\prime}, y^{\prime}\right) \sim p_{g}(x, y)}\left[-\log p_{c}\left(y^{\prime} | x^{\prime}\right)\right].
		\label{equ_cls_fake}
	\end{equation}
	Generator G tries to minimize the loss $\mathcal{L}_{c l s}^{f}$ to produce the class-discriminative maps for generating MR images $x^{\prime}$ that can be classified as the target class $y^{\prime}$.

	\textbf{Cycle consistency loss.} By minimizing the adversarial and classification losses, generator G is trained to generate MR images that are realistic and classified as target class. However, minimizing the losses (Eqs. (\ref{equ_adv})  and Eqs. (\ref{equ_cls_fake})) does not guarantee that the final transformed images preserve the content of input MR  images while changing only the disease-related regions of the input. To alleviate this problem, a forward cycle consistency loss and backward cycle consistency loss \cite{cycle_consistent_loss1,cycle_consistent_loss2} are applied  to the generator. They are  defined as
	\begin{equation}
		\mathcal{L}_{cyc\_tar}  =   \mathbb{E}_{x, y^{\prime}, y}\left[\left\|x_{\text {real}}^{\prime}-\left(G\left(x, y^{\prime}\right)+x\right)\right\|_{1}\right],
	\end{equation}
	\begin{equation}
		\begin{split}
			&\mathcal{L}_{cyc\_org} \\ = &  \mathbb{E}_{x, y^{\prime}, y}\left[\left\|x_{\text {real}}-\left(G\left(x^{\prime}, y\right)+x^{\prime}\right)\right\|_{1}\right]
			\\ =  & \mathbb{E}_{x, y^{\prime}, y}\left[\left\|x_{\text {real}}-\left(G\left((G(x, y^{\prime}) + x), y\right)+x^{\prime}\right)\right\|_{1}\right],
		\end{split}
	\end{equation}where generator G takes in the transformed MR image $x^{\prime}$ and the source class  $y$ as input and tries to reconstruct the MR image $X_{r}= G\left(x^{\prime}, y\right)+x^{\prime}$ of the source domain $y$. The \emph{L}1 norm is adopted as the reconstruction loss.  Note that a single generator is reused twice. The generator is first utilized to transform MR images from the source domain $y$ to MR images of the target domain $y^{\prime}$. Then it is used to reconstruct the MR image of the source domain $y$ from the synthetic MR images of the target domain $y^{\prime}$. For the first utilization, forward cycle consistency loss $\mathcal{L}_{cyc_{-} t a r}$ is adopted. For the second one, backward cycle consistency loss $\mathcal{L}_{cyc_{-} o r g}$ is adopted.

	\textbf{\emph{L}1 penalty.}   The smallest class-discriminative map $\Delta x$ that leads to a real MR image of the target domain $y^{\prime}$ is encouraged. Thus \emph{L}1 penalty is defined as

	\begin{equation}
		\mathcal{L}_{1}(\Delta x)=\|\Delta x\|_{1},
	\end{equation} where $\|\cdot\|_{1}$ is the $\mathbf{\emph{L}}1$ norm.
	
	\textbf{Total Loss.}  The total loss functions to optimize D, C, and G are defined respectively as
	\begin{equation}\mathcal{L}_{D}=-\mathcal{L}_{a d v}, \end{equation}
	\begin{equation}\mathcal{L}_{C}=\mathcal{L}_{c l s}^{r},\end{equation}
	\begin{equation}
		\begin{split}
			\mathcal{L}_{G} = & \mathcal{L}_{a d v}+\lambda_{c l s} \mathcal{L}_{c l s}^{f}+\lambda_{1} \mathcal{L}_{1}(\Delta x)
			\\   & +\lambda_{cyc\_org} \mathcal{L}_{cyc\_org} +\lambda_{cyc\_tar} \mathcal{L}_{cyc\_tar},
			\label{G_loss}
		\end{split}
	\end{equation}
	where $\lambda_{c l s}$,  $\lambda_{1}$ , $\lambda_{cyc\_org}$ and   $\lambda_{cyc\_tar}$ are hyperparameters that control the relative importance of   classification loss, \emph{L}1 penalty, and cycle consistency loss  respectively, compared to the adversarial loss.  $\lambda_{c l s}$ is set as 0.1, $\lambda_{1} $is set as 10,  $\lambda_{cyc\_org}$ is set as 10 and $\lambda_{cyc\_tar}$ is set as 1 empirically throughout the paper. Please note that each loss term is indispensable for the proposed MP-GAN. Without any loss term of the hybrid loss, the training of MP-GAN will become extremely unstable and the learned class-discriminative maps can’t capture the salient features for each pair of classes.
	
	\begin{figure*}
		\centering \includegraphics[width=0.85\linewidth]{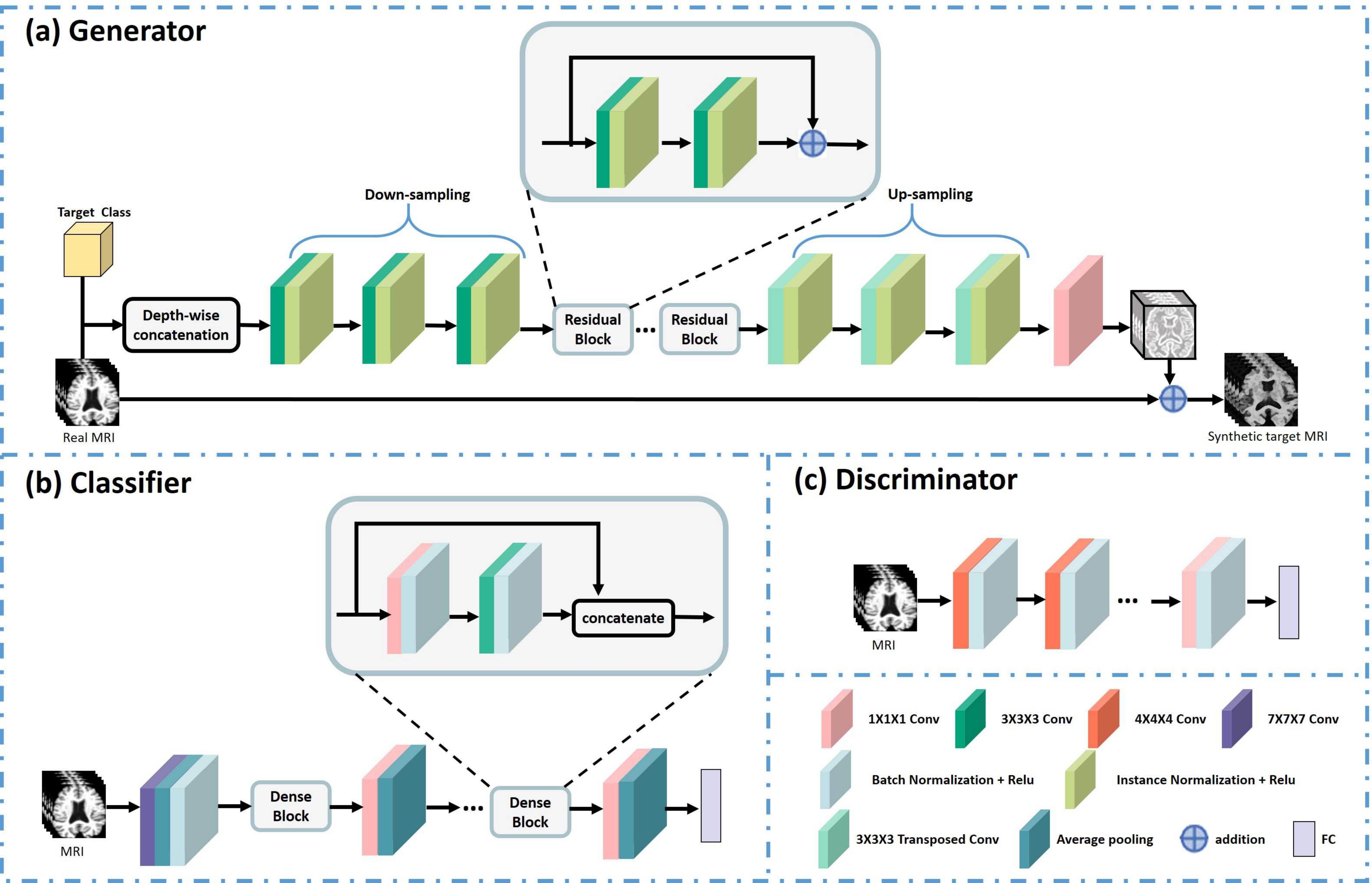}
		\caption{ Network architecture of the proposed MP-GAN. It consists of 3 components:(a) Generator, (b) Classifier, and (c) Discriminator. The generator consists of two convolutional layers for downsampling, three 3D-residual blocks, and two transposed convolutional layers for upsampling. 3D-DenseNet is utilized in the classifier. A standard CNN architecture with 7 convolutional layers with $4 \times 4 \times 4$ and  $1 \times 1 \times 1$ convolutional filters is adopted in the discriminator.}
		\label{fig_FV-GAN-structure}
	\end{figure*}

	During the training process, the following domain settings are defined to train so that all features between any two domains $y$ and $y^{\prime}$ can be visualized for AD analysis.
		\begin{center}
			(1) $y$=\{NC\}, $y^{\prime}$=\{SMC,EMCI,LMCI,AD\},\\
			(2) $y$=\{SMC\}, $y^{\prime}$=\{NC,EMCI,LMCI,AD\},\\
			(3) $y$=\{EMCI\}, $y^{\prime}$=\{SMC,NC,LMCI,AD\},\\
			(4) $y$=\{LMCI\}, $y^{\prime}$=\{SMC,EMCI,NC,AD\},\\
			(5) $y$=\{AD\},  $y^{\prime}$=\{SMC,EMCI,LMCI,NC\}.
		\end{center}
		
		From the algorithm perspective, take the first case (1) as an example, when source domain $y$ is set as NC, the target domain $y^{\prime}$ will be set as one of \{SMC, EMCI, LMCI, AD\}. Please note that all categories in \{SMC, EMCI, LMCI, AD\} set will be trained at least once. In this way, all the above 5 conditions will be trained for MP-GAN, thus a single generator in MP-GAN can learn the class-discriminative maps for each pair of classes, and the salient global features can be captured for multiple classes.   At the testing stage, $y^{\prime}$ is pre-defined according to the requirement of user.  In this paper,  at the testing stage, the morphological features of NC versus all Alzheimer's stages including SMC, EMCI, and LMCI are visualized. MCI is characterized by a slight decline in cognitive abilities. Note that patients with MCI are at increased risk of developing to AD, but do not always do. Thus MCI is significant for morphological feature visualization and further AD analysis.

	The network structure of generator, classifier, and discriminator is shown respectively in Fig. \ref{fig_FV-GAN-structure}.   The network utilized in the generator is ResNet.   3D-ResNet is expanded by adding a spatial dimension to all convolutional and pooling layers in ResNet for the MR image.   By utilizing the shortcut connection, ResNet explicitly reformulates the layers as learning residual functions regarding the input layer, and it transfers feature representations from low layers to the high layers.  More specifically,   assume the target class $y^{\prime}$ is a discrete label, and it is encoded as a one-hot tensor. The target label $y^{\prime}$ is concatenated to the input MRI tensor in a depth-wise manner. Then they are operated by two convolutional layers with a stride size of 2 for downsampling, three 3D-residual blocks\cite{ResNet}, and two transposed convolutional layers with the stride size of 2 for upsampling. In this manner, the target label $y^{\prime}$ is operated with the input MRI tensor of the source label $y$ in each hidden layer of the generator, and the class-discriminative map $\Delta x$ between y and $y^{\prime}$  will be generated. Finally, a synthetic MR image $x^{\prime}$ of target class $y^{\prime}$ is produced by adding the generated class-discriminative map $\Delta x$ and input MRI tensor x.   Instance normalization\cite{IN} is used in all layers except the last output layer for the generator. $3 \times 3 \times 3$ and  $1 \times 1 \times 1$ convolutional filters are employed in generator.  The network utilized in the classifier  is DenseNet \cite{denseNet}.  3D-DenseNet is expanded by adding a spatial dimension to all convolutional and pooling layers in DenseNet for the MR image. Feature-maps learned by all preceding layers are concatenating along the last dimension for the subsequent layers. Through such dense connectivity, feature-maps are reused and the vanishing-gradient problem is alleviated. Meanwhile, 3D-DenseNet can extract discriminative features related to Alzheimer's stage from the whole MR images efficiently. The details of 3D-denseNet can be found in \cite{denseNet,3D-denseNet}. In this paper, the depth is set to 30, the growth rate is set to 12, the number of the Dense-BC block is set to 3, and the reduction is set to 0.5. A standard CNN architecture with 7 convolutional layers with $4 \times 4 \times 4$ and  $1 \times 1 \times 1$ convolutional filters is adopted in the discriminator.  Each convolutional layer is followed by batch normalization\cite{BN} and ReLU.
	
	\begin{table*}
		\caption{Demographic characteristics of the subjects in ADNI dataset.}
		
		\resizebox{\textwidth}{!}{
			
			\begin{tabular}{|l|c|c|c|c|c|c|c|c|c|c|c|c|}
				\hline
				Magnet strength & \multicolumn{8}{c|}{3T} & \multicolumn{4}{c|}{1.5T} \\ \hline
				Source & \multicolumn{2}{c|}{ADNI-1} & ADNI-GO & \multicolumn{5}{c|}{ADNI-2} & \multicolumn{2}{c|}{ADNI-1} & ADNI-GO & ADNI-2 \\ \hline
				Subject & NC & AD & EMCI & NC & SMC & EMCI & LMCI & AD & NC & AD & NC & NC \\ \hline
				Number & 42 & 29 & 142 & 190 & 121 & 309 & 177 & 159 & 171 & 175 & 16 & 81 \\ \hline
				Gender(F/M) & 27 / 15 & 20 / 9 & 67 / 75 & 95 / 95 & 64 / 47 & 139 / 169 & 80 / 97 & 68 / 91 & 89 / 82 & 85 / 90 & 8 / 8 & 45 / 36 \\ \hline
				Age & 76.1$\pm$5.1 & 75.6$\pm$7.9 & 71.7$\pm$7.7 & 74.9$\pm$6.8 & 72.9$\pm$5.6 & 72.3$\pm$7.3 & 72.9$\pm$7.6 & 75.4$\pm$7.9 & 77.7$\pm$5.4 & 76.6$\pm$7.5 & 80.2$\pm$4.8 & 82.5$\pm$4.5 \\ \hline
				Education & 16$\pm$2.8 & 14.7$\pm$2.9 & 15.8$\pm$2.7 & 16.4$\pm$2.7 & 16.8$\pm$2.5 & 16$\pm$2.7 & 16.5$\pm$2.6 & 15.8$\pm$2.7 & 16.0$\pm$2.9 & 14.6$\pm$3.2 & 15.5$\pm$2.5 & 15.9$\pm$2.9 \\ \hline
				MMSE & 29.3$\pm$1.0 & 20.03$\pm$4.8 & 28.2$\pm$1.8 & 28.7$\pm$1.5 & 28.6$\pm$1.7 & 28.0$\pm$2.1 & 26$\pm$3.5 & 20.8$\pm$4.4 & 29.1$\pm$1.2 & 21.5$\pm$4.4 & 29.4$\pm$1.0 & 28.5$\pm$2.6 \\ \hline
				CDR & 0$\pm$0.14 & 1.07$\pm$0.4 & 0.45$\pm$0.19 & 0.07$\pm$0.19 & 0.13$\pm$0.23 & 0.46$\pm$0.22 & 0.58$\pm$0.37 & 0.99$\pm$0.46 & 0$\pm$0.19 & 0.93$\pm$0.49 & 0.07$\pm$0.18 & 0.2$\pm$0.35 \\ \hline
				Samples & 149 & 73 & 471 & 723 & 288 & 1111 & 616 & 501 & 587 & 520 & 72 & 205 \\ \hline
		\end{tabular}}
		\label{table_ADNI}
	\end{table*}
	\section{Experiments and Results}\label{sec_Experiments}
	\subsection{Dataset and Preprocessing}\label{sec_Datasets}
	There are five stages associated with AD progression: Normal Control (NC), Significant memory concern (SMC), early mild cognitive impairment (EMCI), late mild cognitive impairment (LMCI), and Alzheimer's disease(AD). T1-weighted MR images from the Alzheimer's Disease Neuroimaging Initiative (ADNI) public dataset are used for the evaluation purpose. 5316 MR images in ADNI-1, ADNI-go, and ADNI-2 are utilized. It includes 1736 NC subjects, 288 SMC subjects, 1582 EMCI subjects, 616 LMCI subjects, and 1094 AD subjects. Both  1.5T and 3T field strength MR images are used.  Table  \ref{table_ADNI}  lists the demographic characteristics of the subjects.

	ADNI-go and ADNI-2 added 129 and 782 participants respectively to the 819 recruited by ADNI-1\footnote[1]{ http://adni.loni.usc.edu/about/}. In ADNI-1, NC and MCI participants continue to be followed by ADNI-go and ADNI-2. Different from the ADNI-1 dataset,  MCI  is divided into two subtypes, including EMCI, and LMCI in the ADNI-2 dataset. ADNI-go also added a new cohort of people with EMCI, and ADNI-2 added a cohort who were clinically evaluated as cognitively normal but had SMC. Note that SMC is the transitional stage between NC and MCI. The diagnostic criteria are described in the ADNI procedures manual\footnote[2]{http://www.adni-info.org}.

	All MR images are in the neuroimaging informatics technology initiative (NIfTI) format. They are processed using standard operations  in the FSL\footnote[3]{www.fmrib.ox.ac.uk/fsl}  toolbox  \cite{FSL1,FSL2,FSL3} for registering the MR images to MNI space.  The preprocessing pipelines contain three parts: (1) removal of redundant tissues; (2) brain area extraction by BET; (3) linear registration by FLIRT \cite{flirt1,flirt2}.  Lastly,   the T1-MR image is normalized into the range [-1,1], and is fed into the MP-GAN model as a tensor directly without compressing or downsizing.

	\subsection{Experiment Settings}\label{sec-setup}
	The proposed MP-GAN is trained on the ADNI dataset from scratch in an end-to-end manner.  All methods are implemented in TensorFlow\footnote[4]{http://www.tensorflow.org/}.  All experiments are conducted on four NVIDIA GeForce GTX 2080 Ti GPUs.  `Adam'  is utilized as the optimizer for stochastic gradient descent (SGD). The batch size is set to 8 empirically as each MR image is a high-order tensor of $109 \times 91 \times 91$. Since the batch size is relatively small, then the gradients will become unstable, thus there is a need to reduce the learning rate to stabilize the training. According to the experimental results, the learning rate of both generator and classifier is set to 0.001, and the learning rate of the discriminator is set much smaller as  $10^{-4}$.  For evaluation, 80\% of the MR images are allocated for training. The remaining 20\% of the MR images are equally partitioned and used as validation and test data sets respectively. For avoiding bias, the training set, validation set, and test set do not have the MR images from the same subject simultaneously. A single MP-GAN model is trained on a training dataset of all categories, then the morphological features between the source domain and the predefined target domain are visualized on the test set. The validation dataset is utilized to tune hyperparameters to obtain the best model out of several epochs during the training process.
	
	\begin{figure*}
		\centering
		\includegraphics[width=\linewidth]{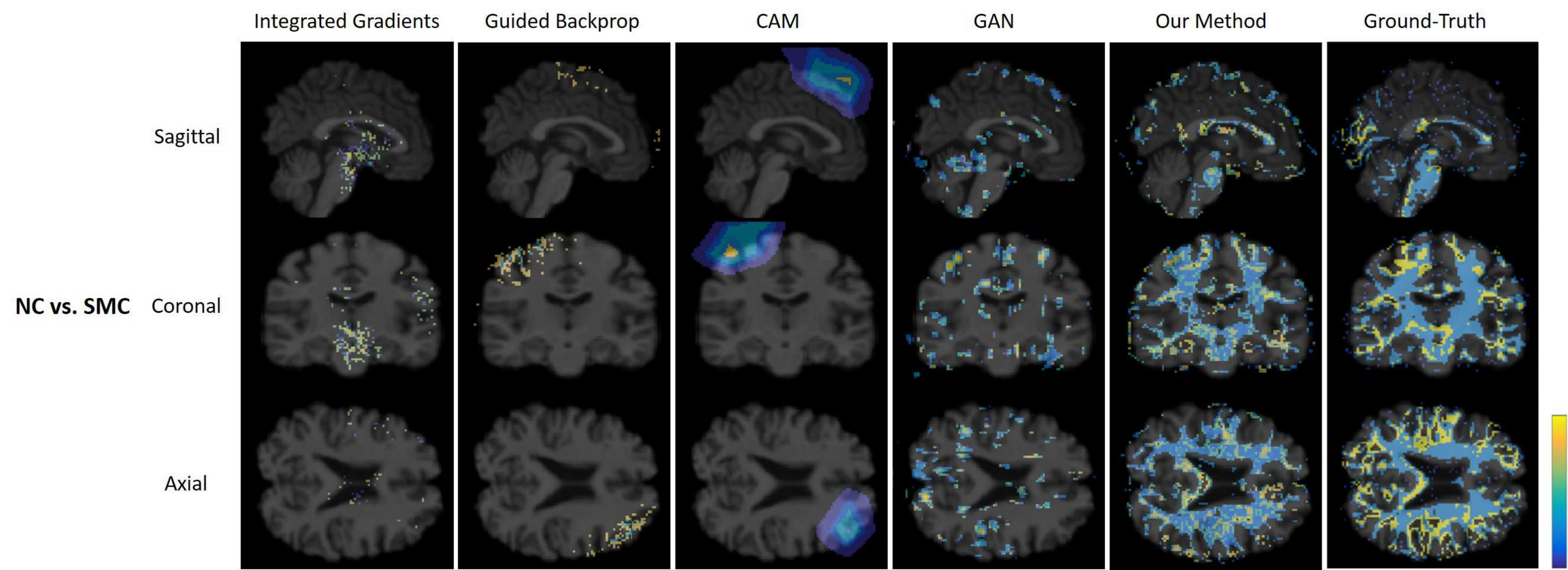}
		\caption{Heatmaps predicted by Integrated Gradients, Guided Backprop, CAM, GAN, and our method are shown in sagittal,coronal, and axial views for NC vs. SMC respectively.}
		\label{fig_ncSMC}
	\end{figure*}
	\begin{figure*}
		\centering
		\includegraphics[width=\linewidth]{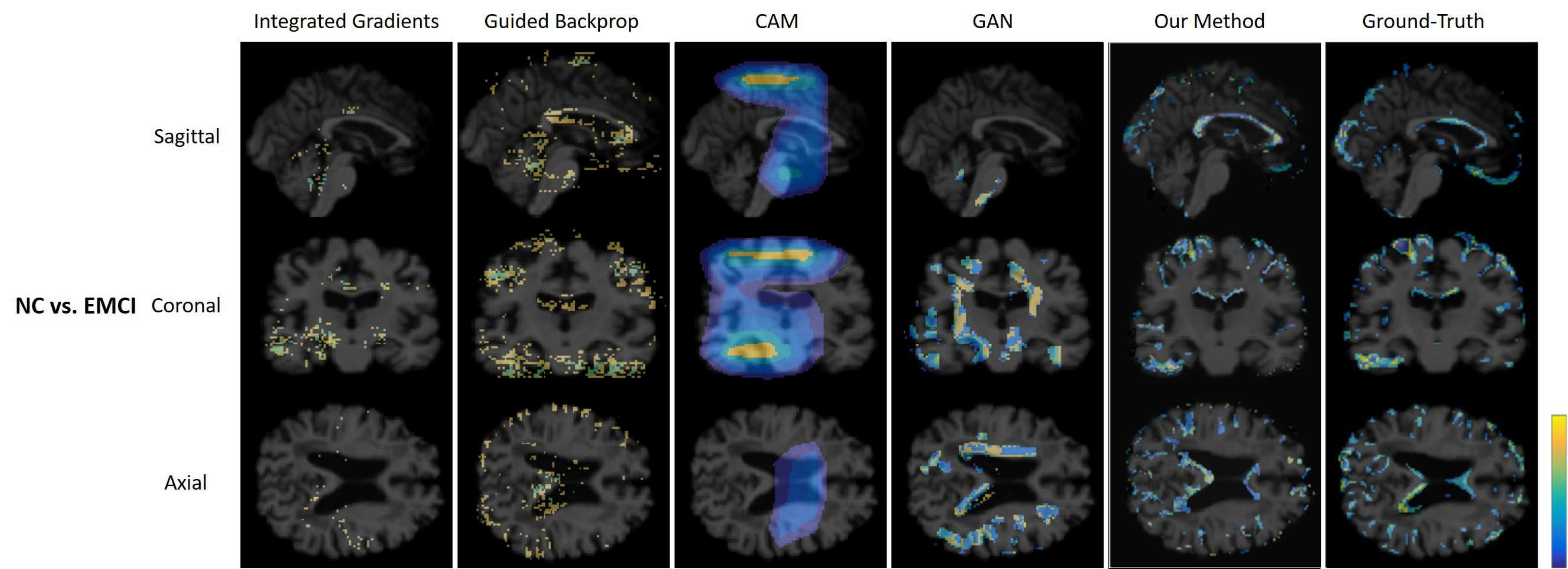}
		\caption{Heatmaps predicted by Integrated Gradients, Guided Backprop, CAM, GAN, and our method are shown in sagittal,coronal,  and axial views for  NC vs. EMCI respectively.}
		\label{fig_ncEMCI}
	\end{figure*}
	\begin{figure*}
		\centering
		\includegraphics[width=\linewidth]{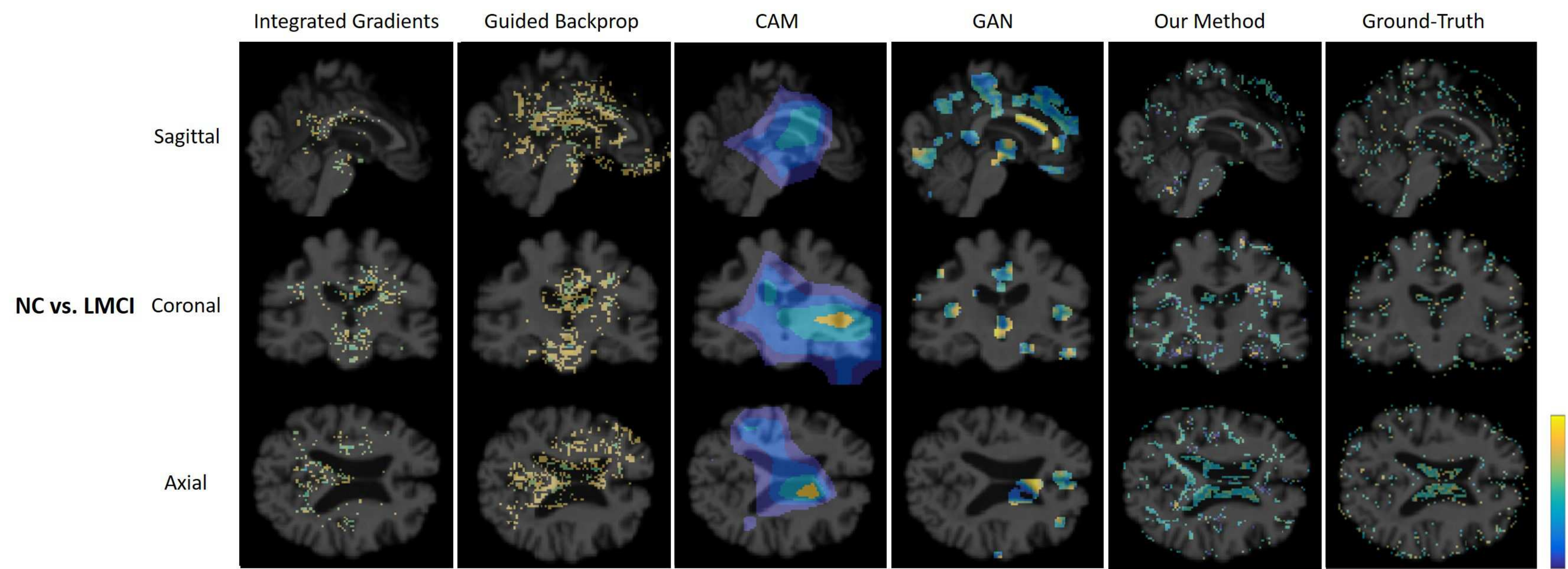}
		\caption{Heatmaps predicted by Integrated Gradients, Guided Backprop, CAM, GAN, and our method are shown in sagittal,coronal,  and axial views for  NC vs. LMCI respectively.}
		\label{fig_ncLMCI}
	\end{figure*}
	\begin{figure*}
		\centering
		\includegraphics[width=\linewidth]{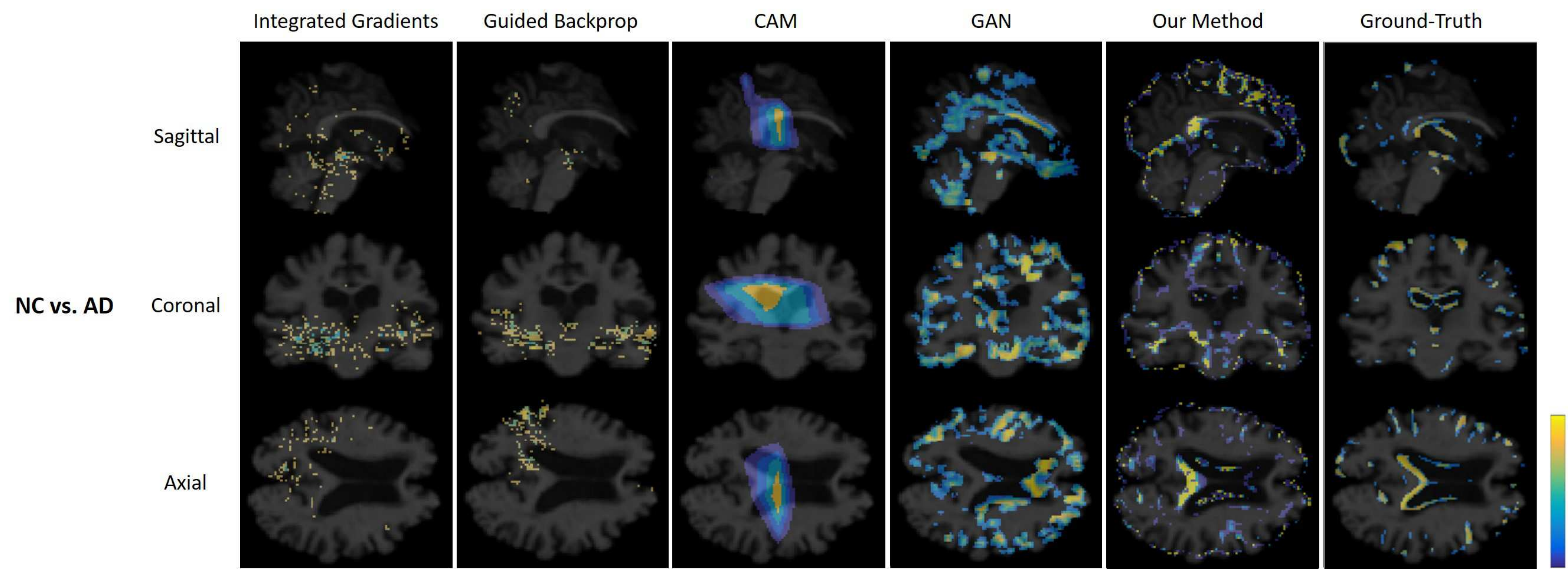}
		\caption{Heatmaps predicted by Integrated Gradients, Guided Backprop, CAM, GAN, and our method are shown in sagittal,coronal,  and axial views for  NC vs. AD respectively.}
		\label{fig_ncAD}
	\end{figure*}
	\subsection{Qualitative Analysis} \label{sec_qualitative}
	In this section,  comprehensive experiments are conducted to show the effectiveness of  MP-GAN.    First, the proposed model is compared with 4 methods: (1)Guided Backpropagation\cite{GuidedBackprop}; (2)Integrated Gradients\cite{IntegratedGradients}; (3)Class Activation Mapping(CAM)\cite{CAM}; and (4)GAN \cite{VAgan}.  For Guided Backpropagation, Integrated Gradients, and CAM, a conventional CNN architecture is used for these networks. More specifically, the CNN architecture consists of 10 convolutional layers followed by the batch normalization and max-pooling layers. After the last convolutional layer, an average pooling layer is utilized instead of the fully connected layer. Besides, for the CAM method, the last layer is designed as described in \cite{CAM} and the last two max-pooling layers are omitted. This allows for more accurate heatmaps due to the higher resolution of the last feature maps. The proposed method is also compared with the conventional GAN \cite{VAgan} to demonstrate our advantages. For a fair comparison, the network structure of the generator and discriminator in GAN is the same as the proposed MP-GAN, and the loss function of GAN is the conventional adversarial loss. The GAN is trained to visualize important regions in MR images between two predefined classes. Furthermore,  the following 4  evaluation groups are set up when compared with the   4 existing methods: (1)NC vs.  SMC, (2)NC vs.  EMCI, (3)NC vs.  LMCI, and (4)NC vs.  AD.  The main reason for this setup is that more meaningful pathological features can be found by comparing with healthy people. It is worth noting that MR images of all 5 classes are trained using only one MP-GAN model, and the class-discriminative map for each evaluation group is visualized at the test stage.   But for the 4 compared methods, one independent binary model is trained for each evaluation group respectively.

	To visually show the quality of heatmaps produced by the proposed model and the 4  existing methods,   one MR image is taken from each evaluation group for qualitative analysis.  From Fig.  \ref{fig_ncSMC} to Fig. \ref{fig_ncAD}, the heatmaps from the   sagittal, coronal,  and axial views are illustrated for each evaluation group respectively. The figures are shown by progression from SMC to AD in order.   From Fig. \ref{fig_ncSMC} to Fig. \ref{fig_ncAD}, it can be seen that the proposed MP-GAN can visualize subtle lesions with contour edge at a finer scale (i.e., voxel-level). More detailed discriminative regions can be depicted, such as the hippocampus, and the corners and boundaries of the ventricle. The highlighted subtle lesions predicted by MP-GAN are relatively more precise than those generated by the other 4 methods. For example, from Fig. \ref{fig_ncEMCI}, it can be observed that the lesions that have much more blurred boundary and are difficult to recognize can be delineated by  MP-GAN.    More specifically, the corpus callosum with irregular sulcus is depicted accurately by  MP-GAN from the sagittal-view and coronal-view in Fig. \ref{fig_ncEMCI}.  Atrophy of the corpus callosum may lead to functional disability because of reduced interhemispheric integration. It is a region that has been examined intensively for indications of EMCI \cite{ccAD}.    On the other hand, Integrated Gradients and Guided Backprop tend to focus on some small parts of the lesions rather than the whole lesions. Because some subtle voxels of the lesion might be more salient than the other voxels of the whole lesion.  This proves that the feature visualization methods based on classification only focus on the most discriminative features and ignore the rest. It is difficult to interpret the results produced by CAM due to the low-resolution.  Moreover, the regions visualized by GAN  seem to cover parts of ground-truth affected by the AD for NC vs. AD as shown in Fig. \ref{fig_ncAD}.  However,  they are not close to ground-truth, this is because the training of GAN is unstable. In summary, the results of the proposed MP-GAN are closer to the ground-truth compared with the other 4 existing methods. This implies that MP-GAN can benefit from the multidirectional mapping mechanism and the hybrid loss function. MP-GAN is more sensitive to subtle structural changes in MR images caused by cognitive decline.

	The ADNI diagnostic criteria for each Alzheimer's stage are briefly described as following.   (1) NC participants have no subjective or informant-based complaints of memory decline, and they have a normal cognitive performance; (2) SMC participants have subjective memory concerns assessed by the Cognitive Change Index(CCI). They have no informant-based complaint of memory impairment or decline, and they have a normal cognitive performance on the Wechsler Logical Memory Delayed Recall (LM-delayed) and the Mini-Mental State Examination (MMSE)\cite{MMSE}; (3) EMCI participants have a subtle cognitive decline. Their abnormal memory function is approximately 1 standard deviation below normative performance, and their MMSE total score is greater than 24; (4) LMCI participants have a memory concern. Clinical Dementia Rating (CDR) of LMCI participants is 0.5, and the Memory Box (MB) score must be at least 0.5; (5) AD participants have a significant memory concern. The MMSE score of AD participants is between 20 and 26, and CDR is  0.5 or 1.0.

\begin{figure*}
		\centering
		\includegraphics[width=\linewidth]{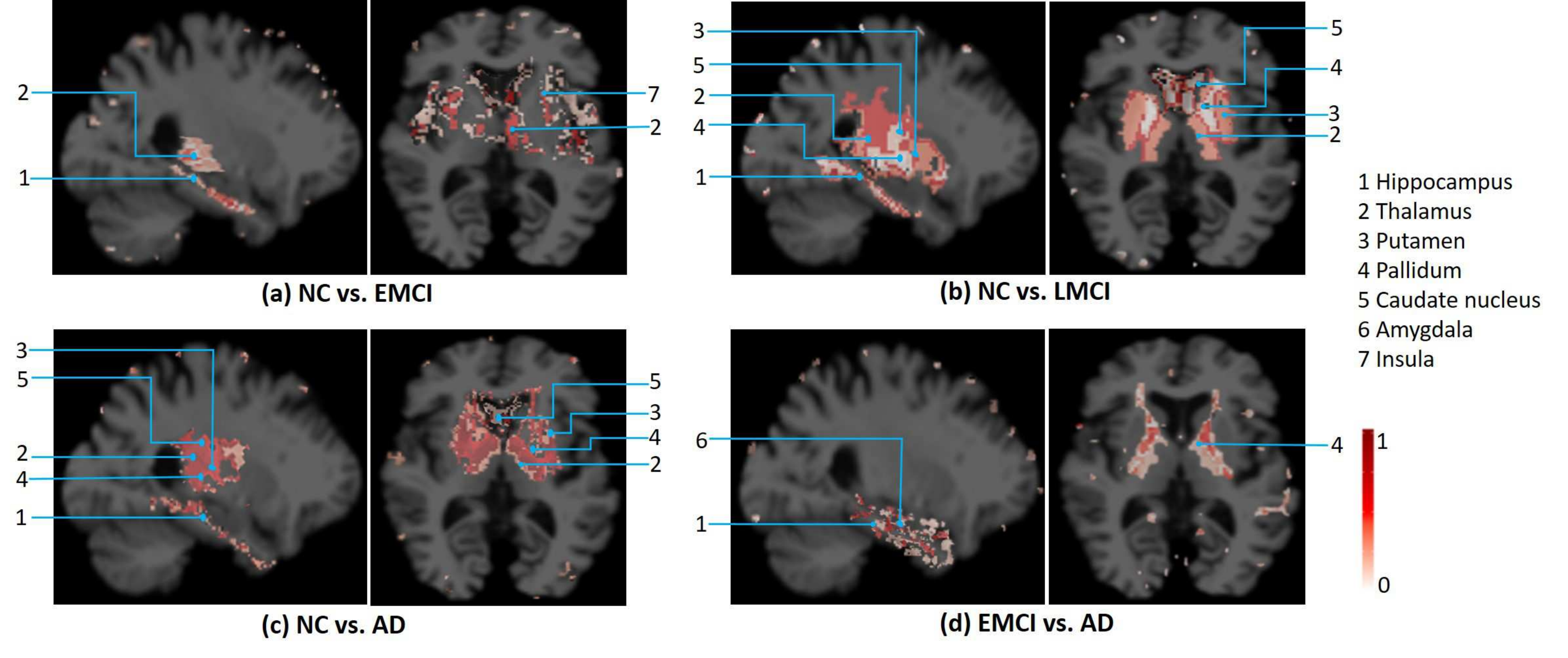}
		\caption{Class-discriminative maps generated by  MP-GAN are shown as a colored overlay over the MR images.  The regions affected by the progression of AD are reliably captured by MP-GAN for four evaluation groups respectively.}	
		\label{fig_annotation}
	\end{figure*}

	To further analyze the visualization results of the proposed MP-GAN from a clinical perspective,  the two-view slices in another coordinate of (33,55,39)  are shown in Fig. \ref{fig_annotation}. Note that the three-view slices shown from Fig. \ref{fig_ncSMC} to Fig. \ref{fig_ncAD} are in the coordinate of (44,55,47). From Fig. \ref{fig_annotation},  the following observations can be made.  (1) For all four evaluation groups, the proposed MP-GAN can delineate the discriminative lesions clearly. More specifically,  lesions visualized by MP-GAN are hippocampus, thalamus, putamen, pallidum, caudate nucleus, amygdala, and insula\cite{ADpriorRegion, MLvisualResults, classificationVisualResults}.  It is worth noting that the discriminative capability of these brain regions in clinical analysis has already been validated by previous studies\cite{ADpriorRegion1, ADpriorRegion2, ADregion1, ADregion2}.  This implies the feasibility of the proposed MP-GAN;  (2) The morphological changes including global atrophy (e. g. smaller volumes of hippocampus or amygdala) and shape changes are visualized by class-discriminative map (indicated by color). These morphological changes are related to AD disease progression and cognitive decline severity;  (3) For 4 evaluation groups, identified multiple regions are overlapped or localized at similar brain regions.  For instance,  the regions of NC vs. LMCI and NC vs. AD are similar because LMCI might develop to AD. Meanwhile, since the features between LMCI and AD are very subtle, thus some visualized regions of NC vs. LMCI and NC vs. AD are overlapped, but the atrophy severity of each lesion is different (indicated by color). The lesions visualized for EMCI vs. AD  and NC vs. AD also have some common regions, such as the hippocampus  and pallidum.   Furthermore,  it is reasonable that the overlap regions between NC vs. EMCI and NC vs. AD might not be identified for EMCI vs. AD, and some regions such as the amygdala which are specific to EMCI vs. AD can be identified;  (4)  Along with the progression from EMCI to AD,  from Fig. \ref{fig_annotation}(a) to Fig. \ref{fig_annotation}(c),  it can be observed that the intensity values (i.e., light salmon color) in the heatmaps are gradually increased (i.e., change to crimson) at various brain locations, and some of them are accumulated at the  annotated regions. These results suggest that the class-discriminative maps generated by the proposed MP-GAN have the potential to provide some extra information regarding the AD progression, and it may reveal the gradual atrophic process of the human brain due to cognitive decline. Furthermore, the severity of cognitive decline is also reflected in ADNI diagnostic criteria for each Alzheimer's stage as described above. In summary, the above observations imply the robustness of MP-GAN in visualizing morphological features for different Alzheimer's stages.
	\begin{figure*}
		\centering \includegraphics[width=0.8\linewidth]{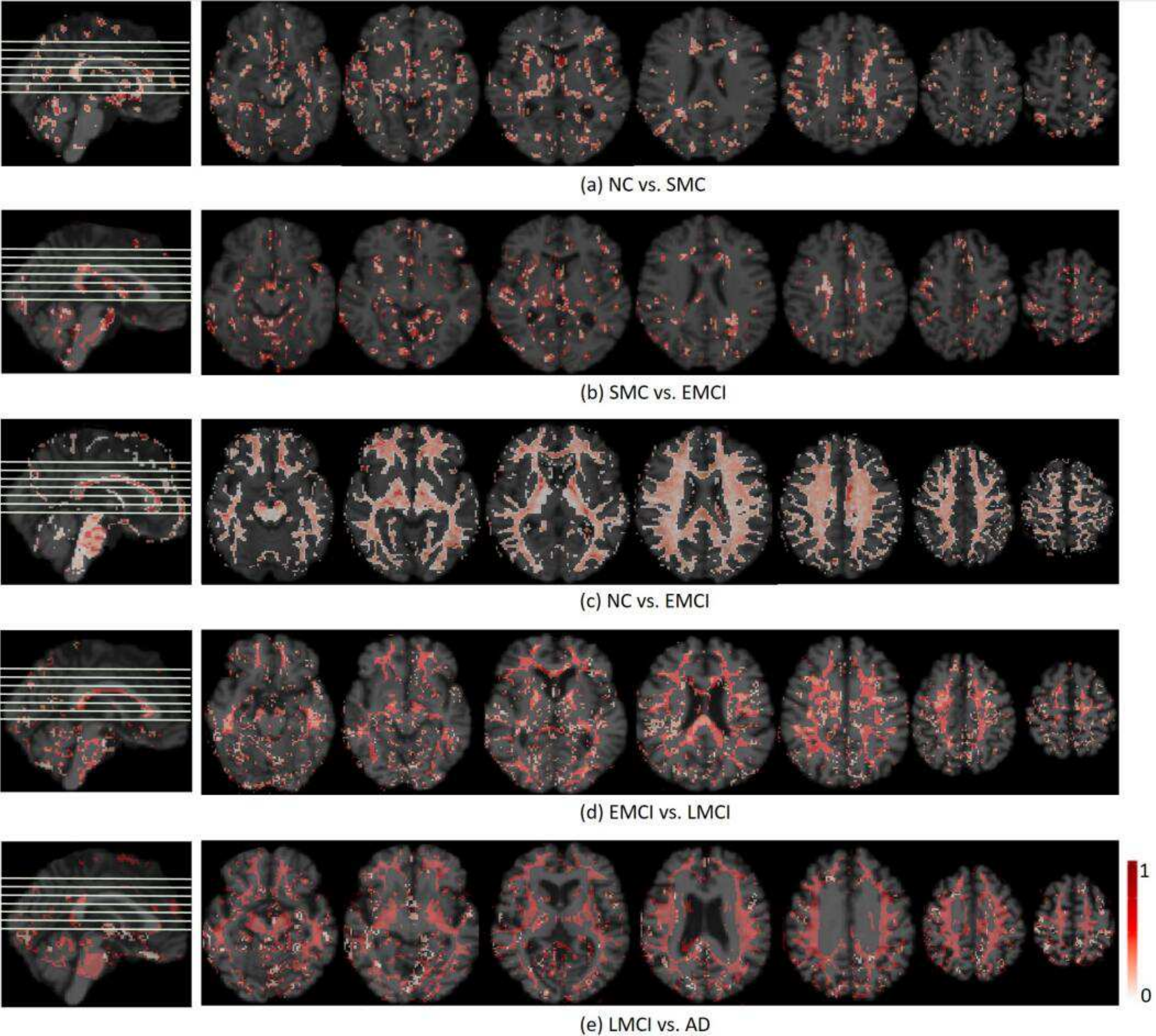}
		\caption{Distribution of  class-discriminative maps visualized by MP-GAN for five evaluation groups respectively.}
		\label{fig_multislice}
	\end{figure*}

	For further visualization analysis, 5 evaluation groups are investigated respectively in Fig. \ref{fig_multislice}.  The results show that the important brain regions visualized by the proposed method are consistent with regions in Fig. \ref{fig_annotation}.   More specifically,   by aligning the automatic anatomical labeling (AAL) map with the class-discriminative maps visualized in Fig. \ref{fig_multislice}, each region in the class-discriminative map will be matched to the specific ROI index and name in AAL. The disease-related regions visualized by MP-GAN are listed in Table \ref{table_list_region}.  Note that the suffix `L' denotes the left brain, and the suffix `R'  denotes the right brain.  The following observations  can be made from Fig. \ref{fig_multislice} and Table \ref{table_list_region}. (1) The brain regions visualized by the proposed method for NC vs. SMC  are precental gyrus, middle frontal gyrus, inferior frontal gyrus, median cingulate, paracingulate gyri, parahippocampal gyrus, superior occipital gyrus, postcentral gyrus and thalamus; (2) The brain regions visualized by the proposed method for SMC vs.  EMCI  are rolandic operculum,  insula,  parahippocampal gyrus, amygdala, superior occipital gyrus, middle occipital gyrus,  postcentral gyrus, superior parietal gyrus and precuneus;  (3) The brain regions visualized by the proposed method for NC vs. EMCI  are the middle frontal gyrus,  posterior cingulate gyrus, calcarine fissure and surrounding cortex, cuneus, superior occipital gyrus, fusiform gyrus,    postcentral gyrus, lenticular nucleus, putamen and inferior temporal gyrus;  (4) The brain regions visualized by the proposed method for EMCI vs. LMCI are the superior frontal gyrus, orbital part, inferior frontal gyrus,  opercular part,   hippocampus, parahippocampal gyrus, calcarine fissure and surrounding cortex, lingual gyrus, inferior occipital gyrus and  fusiform gyrus;  (5) The brain regions visualized by the proposed method for LMCI vs.  AD  are the middle frontal gyrus, orbital part, inferior frontal gyrus, triangular part, hippocampus, calcarine fissure and surrounding cortex,  lingual gyrus, middle occipital gyrus, precuneus, lenticular nucleus and putamen.  These regions also agree with the existing research findings. To sum up, the  lesions visualized by the proposed model are highly suggestive and effective for tracking the progression of AD.
	\begin{figure}
		\centering \includegraphics[width=\linewidth]{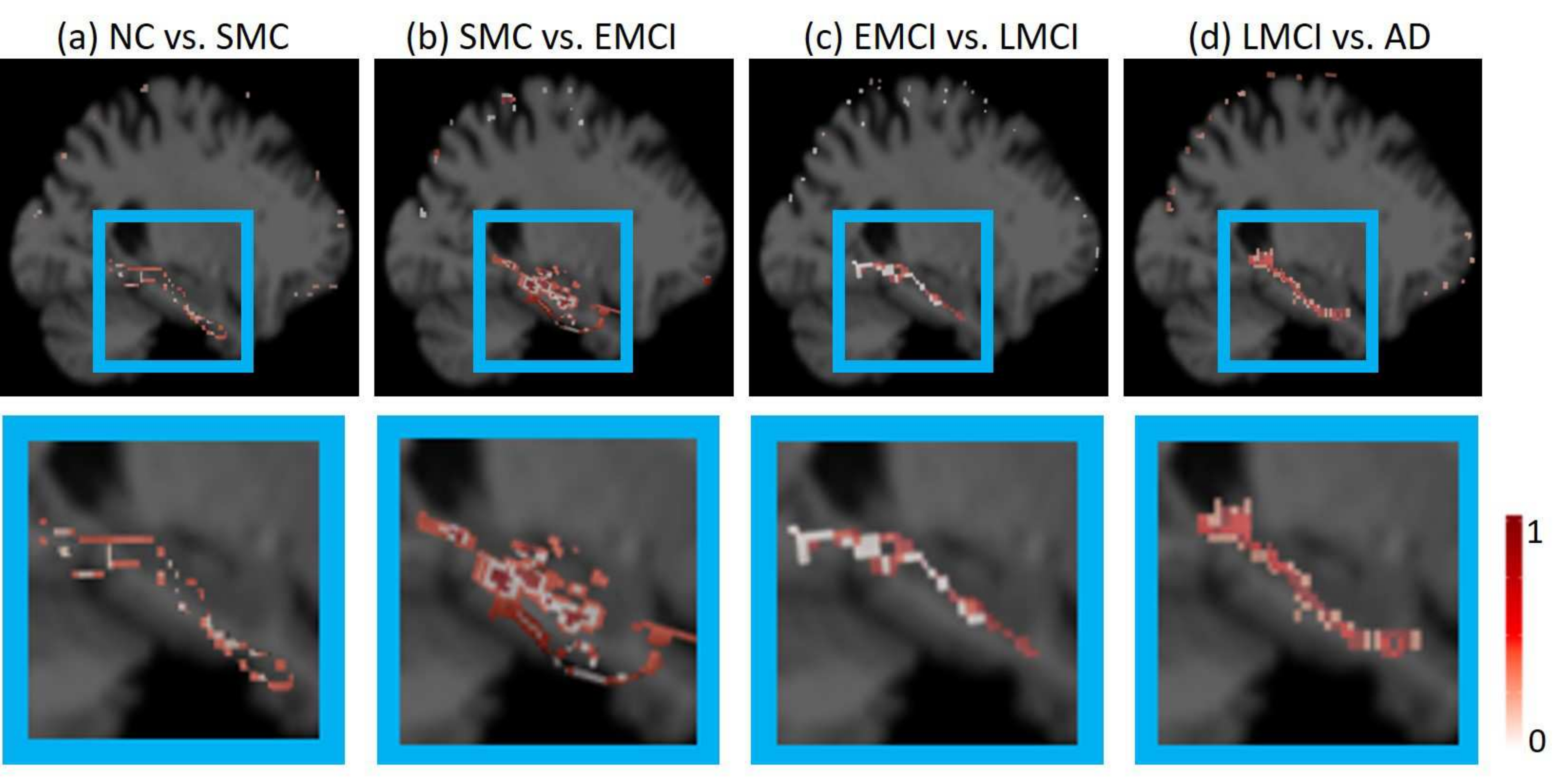}
		\caption{Visualization results of hippocampus by MP-GAN in sagittal view and corresponding zoomed regions. The subfigures at the bottom are the zoom of the original subfigures for better observation.}
		\label{fig_enlarge1}
	\end{figure}
	\begin{table*}
		\center
		\caption{Indices and names of regions visualized by MP-GAN using the AAL template.}
		
		\resizebox{\textwidth}{!}{
			\begin{tabular}{|c|c|c|c|c|c|c|c|c|c|}
				\hline
				\multicolumn{2}{|c|}{NC vs. SMC} & \multicolumn{2}{c|}{SMC vs. EMCI} & \multicolumn{2}{c|}{NC vs. EMCI} & \multicolumn{2}{c|}{EMCI vs. LMCI} & \multicolumn{2}{c|}{LMCI vs. AD} \\ \hline
				ROI index & ROI name & ROI index & ROI name  & ROI index & ROI name & ROI  index & ROI name & ROI index & ROI name  \\ \hline
				1 & Precentral\_L & 17 & Rolandic\_Oper\_L & 7 & Frontal\_Mid\_L & 5 & Frontal\_Sup\_Orb\_L & 9 & Frontal\_Mid\_Orb\_L \\ \hline
				8 & Frontal\_Mid\_R & 29 & Insula\_L & 36 & Cingulum\_Post\_R & 12 & Frontal\_Inf\_Oper\_R & 14 & Frontal\_Inf\_Tri\_R \\ \hline
				10 & Frontal\_Mid\_Orb\_R & 39 & ParaHippocampal\_L & 43 & Calcarine\_L & 37 & Hippocampus\_L & 37 & Hippocampus\_L \\ \hline
				12 & Frontal\_Inf\_Oper\_R & 42 & Amygdala\_R & 45 & Cuneus\_L & 39 & ParaHippocampal\_L & 38 & Hippocampus\_R \\ \hline
				34 & Cingulum\_Mid\_R & 49 & Occipital\_Sup\_L & 46 & Cuneus\_R, & 40 & ParaHippocampal\_R & 43 & Calcarine\_L \\ \hline
				39 & ParaHippocampal\_L & 51 & Occipital\_Mid\_L & 50 & Occipital\_Sup\_R & 43 & Calcarine\_L & 48 & Lingual\_R \\ \hline
				40 & Parahippocampal\_R & 52 & Occipital\_Mid\_R & 56 & Fusiform\_R & 47 & Lingual\_L & 52 & Occipital\_Mid\_R \\ \hline
				50 & Occipital\_Sup\_R & 58 & Postcentral\_R & 57 & Postcentral\_L & 50 & Occipital\_Sup\_R & 67 & Precuneus\_L \\ \hline
				57 & Postcentral\_L & 60 & Parietal\_sup\_R & 74 & Putamen\_R & 54 & Occipital\_Inf\_R & 68 & Precuneus\_R \\ \hline
				78 & Thalamus\_R & 67 & Precuneus\_L & 90 & Temporal\_Inf\_R & 55 & Fusiform\_L & 74 & Putamen\_R \\ \hline
		\end{tabular}}
		\label{table_list_region}
	\end{table*}

	The performance of  MP-GAN to visualize the subtle lesions in the hippocampus is further investigated.  The class-discriminative maps of the hippocampus in the sagittal view are visualized in Fig. \ref{fig_enlarge1}.  Specifically, the following 4 neighborhood evaluation groups are further explored: (a) NC vs. SMC; (b) SMC vs. EMCI; (c) EMCI vs. LMCI; (d) LMCI vs. AD.  From Fig. \ref{fig_enlarge1}, it can be observed that the zoomed regions preserve more details in the hippocampus. In particular,  in the earlier stages of AD such as (a) NC vs. SMC and (b) SMC vs. EMCI, the visualized lesions are extremely subtle and scattered around the boundary of the hippocampus.  In the later stages of AD such as (c) EMCI vs. LMCI and (d) LMCI vs. AD, the visualized lesions are accumulated at the core region of the hippocampus.  Furthermore,  Fig. \ref{fig_enlarge1}(a) to Fig. \ref{fig_enlarge1}(d) reflect the shape change and atrophy of the hippocampus qualitatively as the progressive  deterioration from SMC to AD.   It has already been validated by the previous studies  \cite{biomarkerNew} that the hippocampus is significant for identifying biomarkers in clinical practice.  Although the volume loss and shape change of the hippocampus can not be quantitatively measured in this work, the visualized lesions of the hippocampus are beneficial for identifying the biomarkers in future work. Based on these visualized lesions  in Fig. \ref{fig_enlarge1}, the existing biomarkers such as Brain boundary shift integral (BBSI)\cite{BBSI},  Scoring by Non-local Image Patch Estimator(SNIPE)\cite{SNIPE}, and other grading biomarkers\cite{gradingBiomarker} can be computed. Furthermore, new potential biomarkers reflecting the shape change and brain atrophy might be discovered based on these visualized lesions in the hippocampus in future work.

	\subsection{Quantitative Analysis}\label{sec_quantitative}

	In this section, the following 4 metrics are computed to assess visual quality.  (1) Normalized Cross-Correlation (NCC). NCC \cite{VAgan} is calculated between the ground-truth maps and the predicted class-discriminative maps.  The higher the NCC, the more correlation between ground-truth maps and the predicted class-discriminative maps. For Integrated Gradients, Guided Backprop, and CAM,  the visualized heatmaps for predicting positive class are utilized to calculate the NCC;    (2) Peak Signal-to-Noise Ratio (PSNR).  PSNR \cite{PSNR} is also calculated between the ground-truth maps and the predicted class-discriminative maps on the test data set.  Similar to NCC,  the higher the PSNR, the closer between ground-truth maps and the predicted class-discriminative maps;  (3) Structural Similarity Index Measure (SSIM) \cite{SSIM}.  Different from NCC and PSNR, SSIM in each iteration is calculated between synthetic images and real images on the validation dataset.  Higher SSIM indicates better reconstructed MR image quality.  By computing SSIM in each iteration, the convergency of the model is further validated;  (4) Classification metrics \cite{ADx1} such as AUC, ACC, Sensitivity, and Specificity for data augmentation.  Note that the purpose of  SSIM and the classification metrics is to demonstrate that the proposed MP-GAN  can generate images close to real distribution, thus it validates that MP-GAN can capture salient global features in class-discriminative maps. For  NCC and PSNR, the 4 existing methods are compared. For SSIM, only GAN is compared since the other 3 methods are based on classification. Similarly, for the classification metrics, only  GAN is compared since the classification performance is based on synthetic data augmentation by the proposed MP-GAN and GAN.
	
	\begin{figure}
		\centering
		\includegraphics[width=0.95\linewidth]{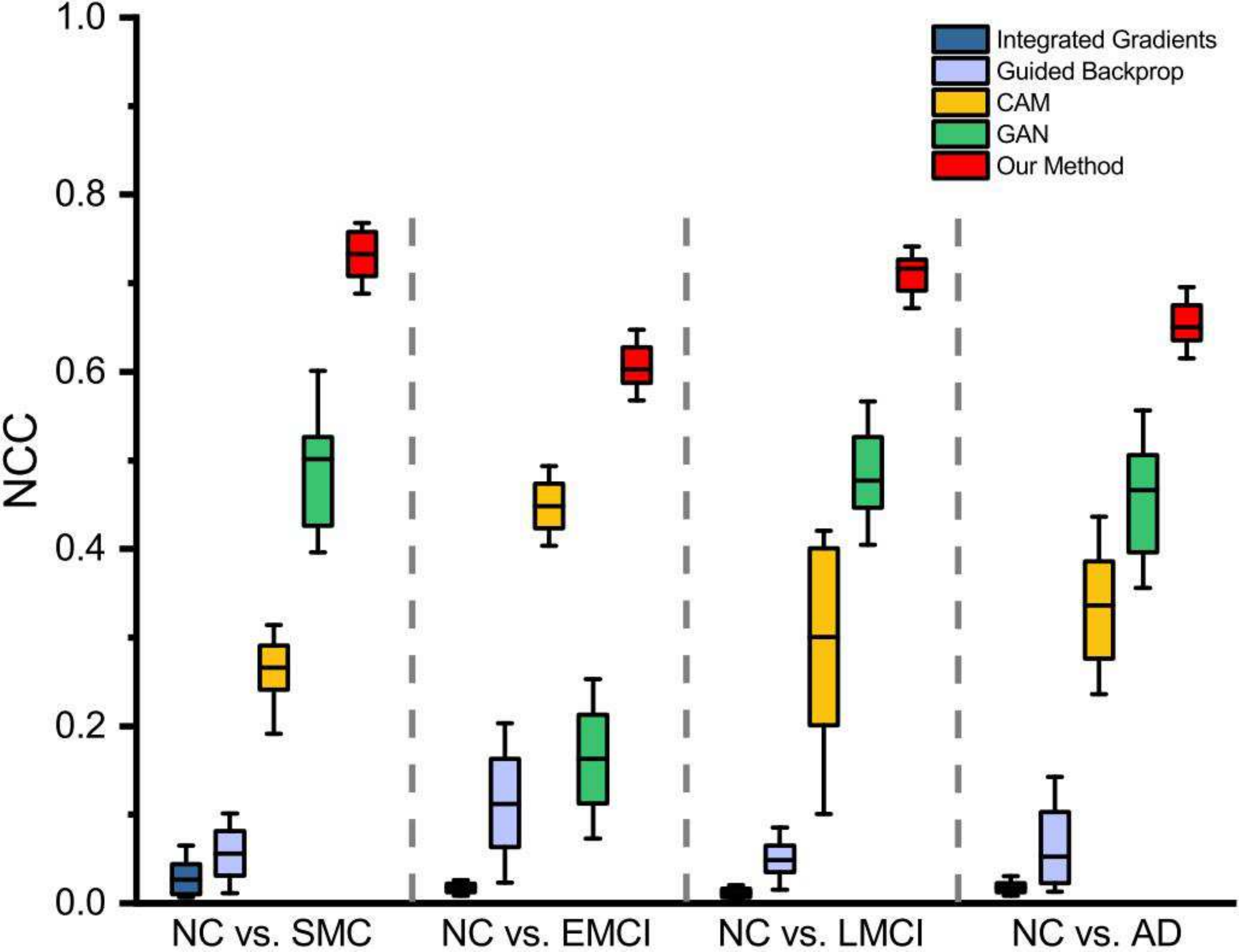}
		\caption{Box-plots of NCC for different models.}
		\label{fig_NCC}
	\end{figure}
	
	The NCC results shown in Fig. \ref{fig_NCC} are mostly consistent with the qualitative results shown from Fig. \ref{fig_ncSMC} to Fig. \ref{fig_ncAD}.   The proposed MP-GAN achieves significantly higher NCC than the other 4 existing methods. It indicates that the distribution of class-discriminative maps generated by MP-GAN is the closest to ground-truth maps. The three methods based on classification (Integrated Gradients, Guided Backprop, and CAM) achieve a low NCC score due to their exclusive focus on local features.  GAN performs better than 3 classification-based feature visualization methods for NC vs. SMC, NC vs. LMCI, and NC vs. AD. This implies that the GAN architecture can capture global features, which alleviates the limitations of feature visualization methods based on classification.  Above all, the proposed MP-GAN achieves the highest correlation scores compared with the other 4 existing methods in all 4 evaluation groups.
	
	\begin{figure}
		\centering
		\includegraphics[width=0.95\linewidth]{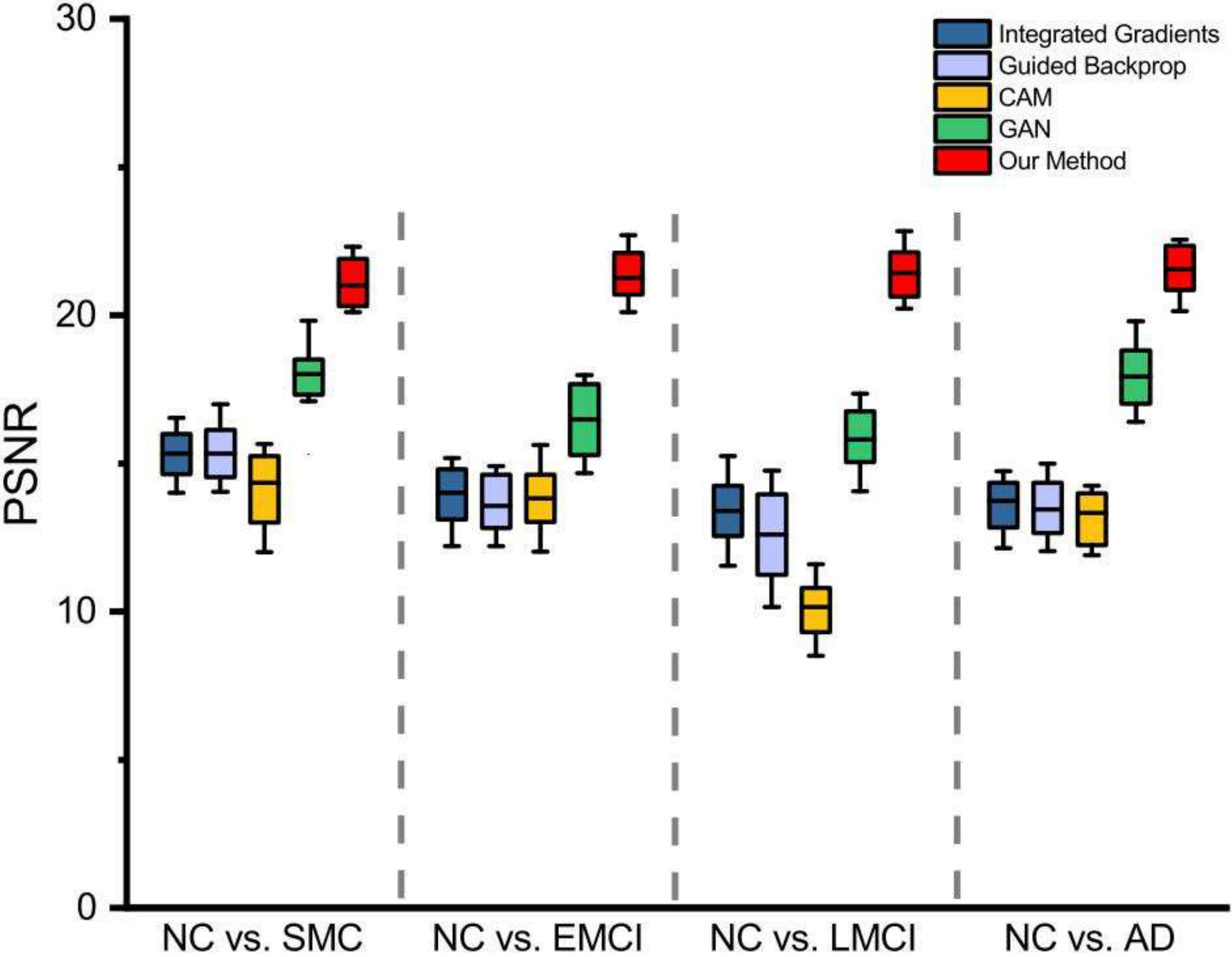}
		\caption{Box-plots of PSNR for different models.}
		\label{fig_PSNR}
	\end{figure}

	From Fig. \ref{fig_PSNR}, it can be seen that the proposed MP-GAN achieves the best PSNR compared with the other 4 existing methods. This is also consistent with NCC results in Fig. \ref{fig_NCC} and the qualitative results shown from Fig. \ref{fig_ncSMC} to Fig. \ref{fig_ncAD}.  The class-discriminative maps visualized by MP-GAN are closer to ground-truth. This is because MP-GAN benefits from the multidirectional mapping mechanism and the hybrid loss function. Meanwhile, MP-GAN can be trained on MR images of all classes with only one model. In this manner, the common features unrelated to the disease can be reused, thus all salient global features can be captured in class-discriminative maps for different Alzheimer's stages.

	\begin{figure}
		\centering
		\includegraphics[width=0.95\linewidth]{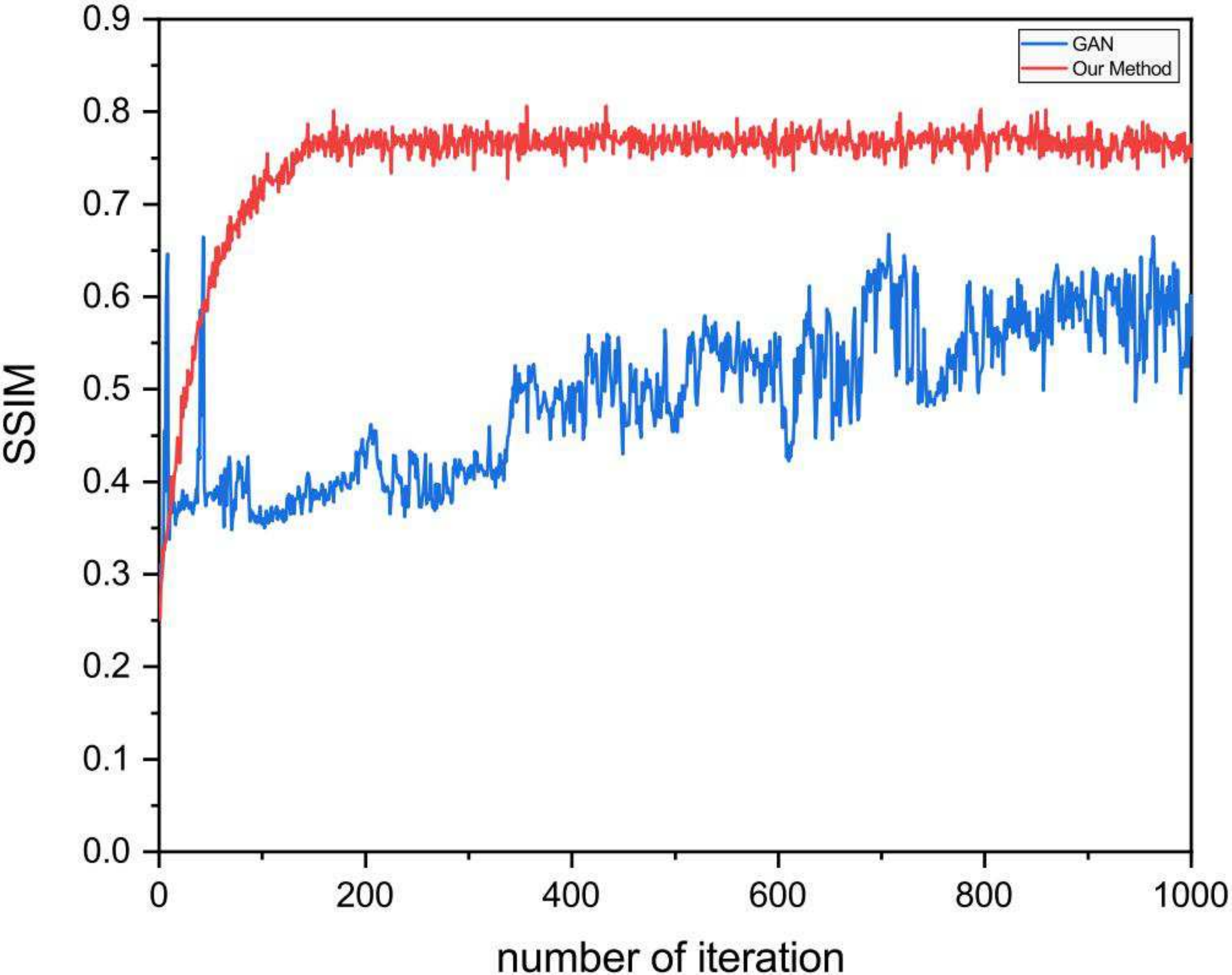}
		\caption{Convergence curves for NC vs. SMC. }
		\label{fig_SSIMncSMC}
	\end{figure}
	
	\begin{figure}
		\centering
		\includegraphics[width=0.95\linewidth]{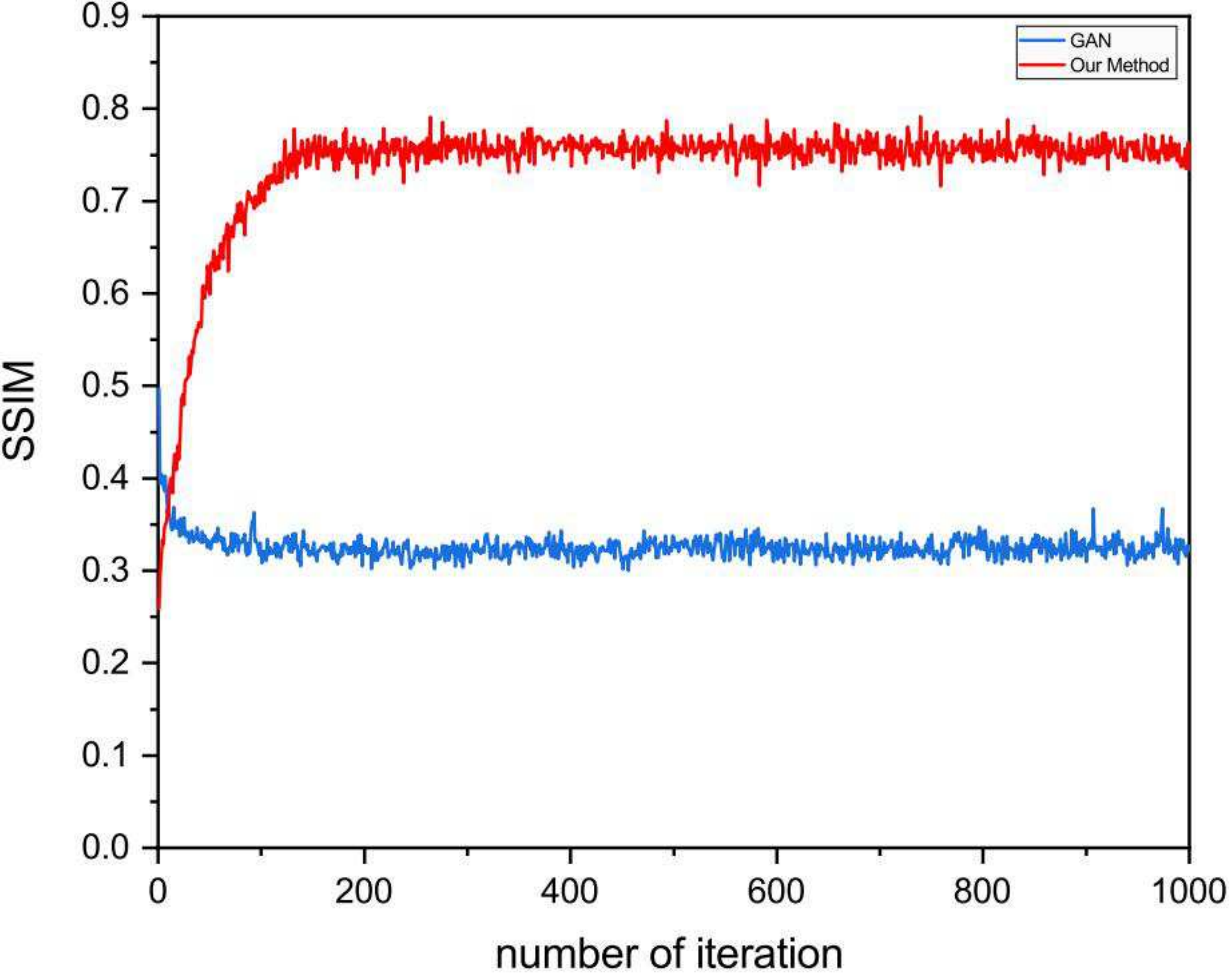}
		\caption{Convergence curves  for NC vs. EMCI.	}
		\label{fig_SSIMncEMCI}
	\end{figure}
	
	\begin{figure}
		\centering
		\includegraphics[width=0.95\linewidth]{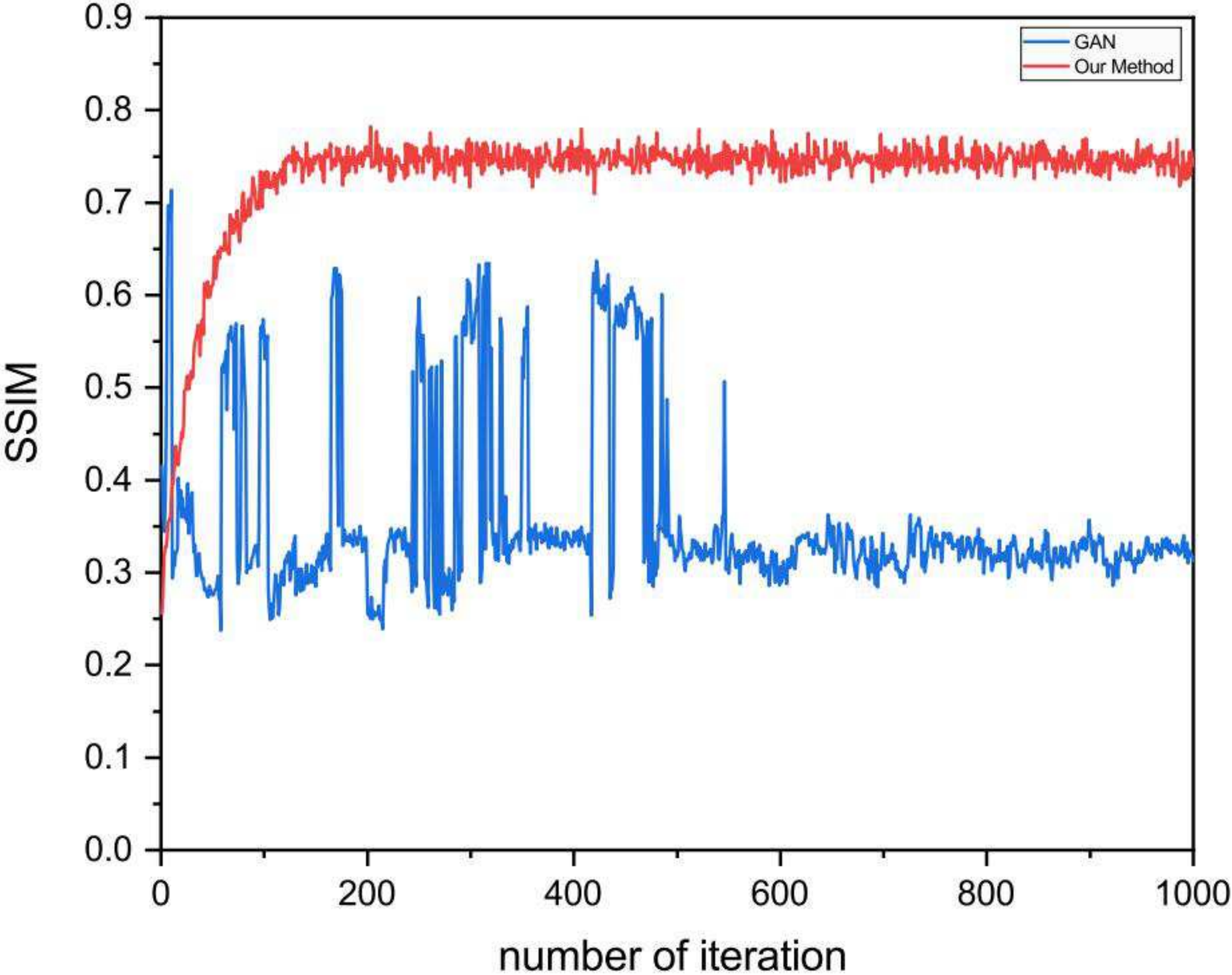}
		\caption{Convergence curves for NC vs. LMCI.}
		\label{fig_SSIMncLMCI}
	\end{figure}
	\begin{figure}
		\centering
		\includegraphics[width=0.95\linewidth]{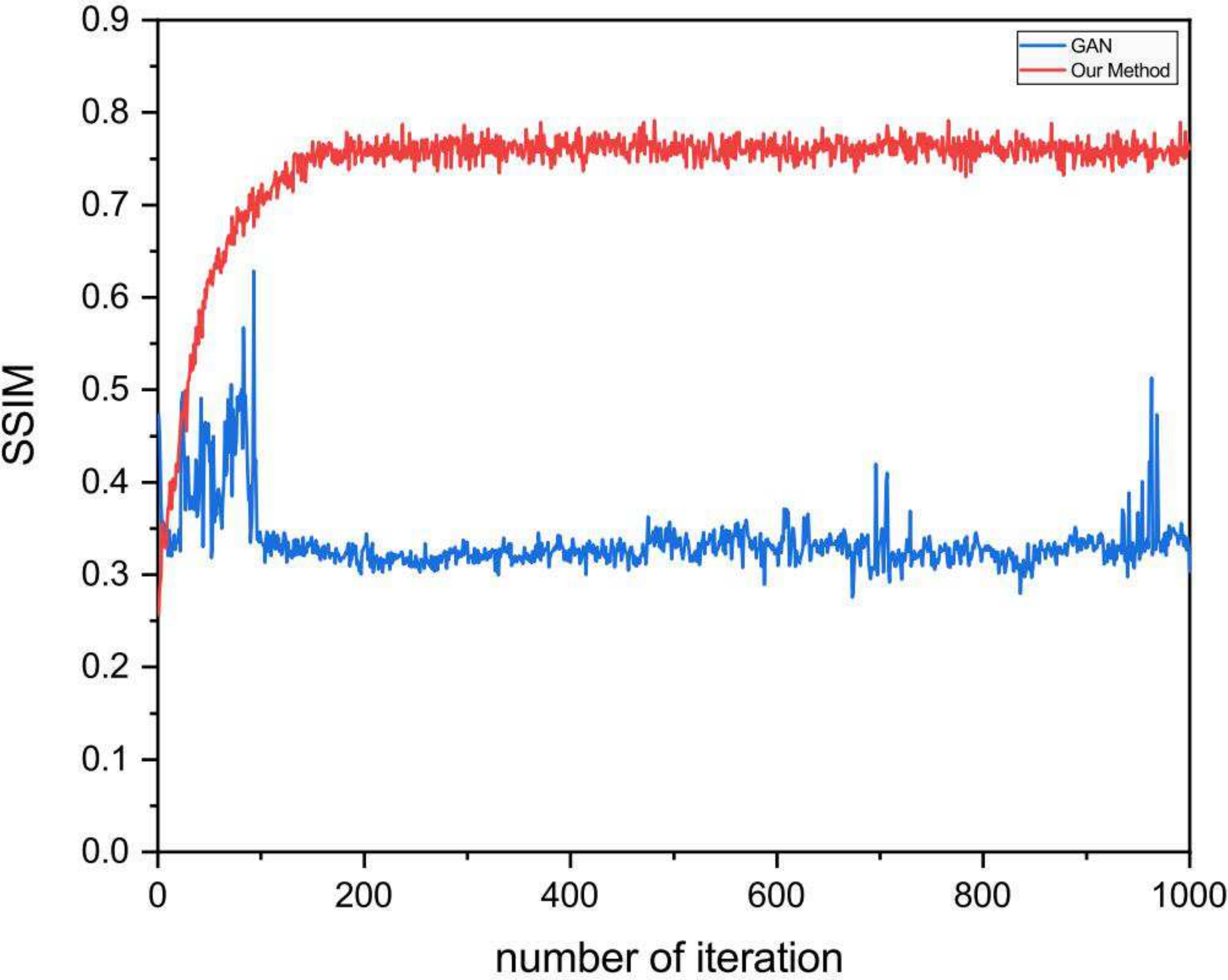}
		\caption{Convergence curves  for NC vs. AD.}
		\label{fig_SSIMncAD}
	\end{figure}

	\begin{figure}
		\centering
		\includegraphics[width=0.9\linewidth]{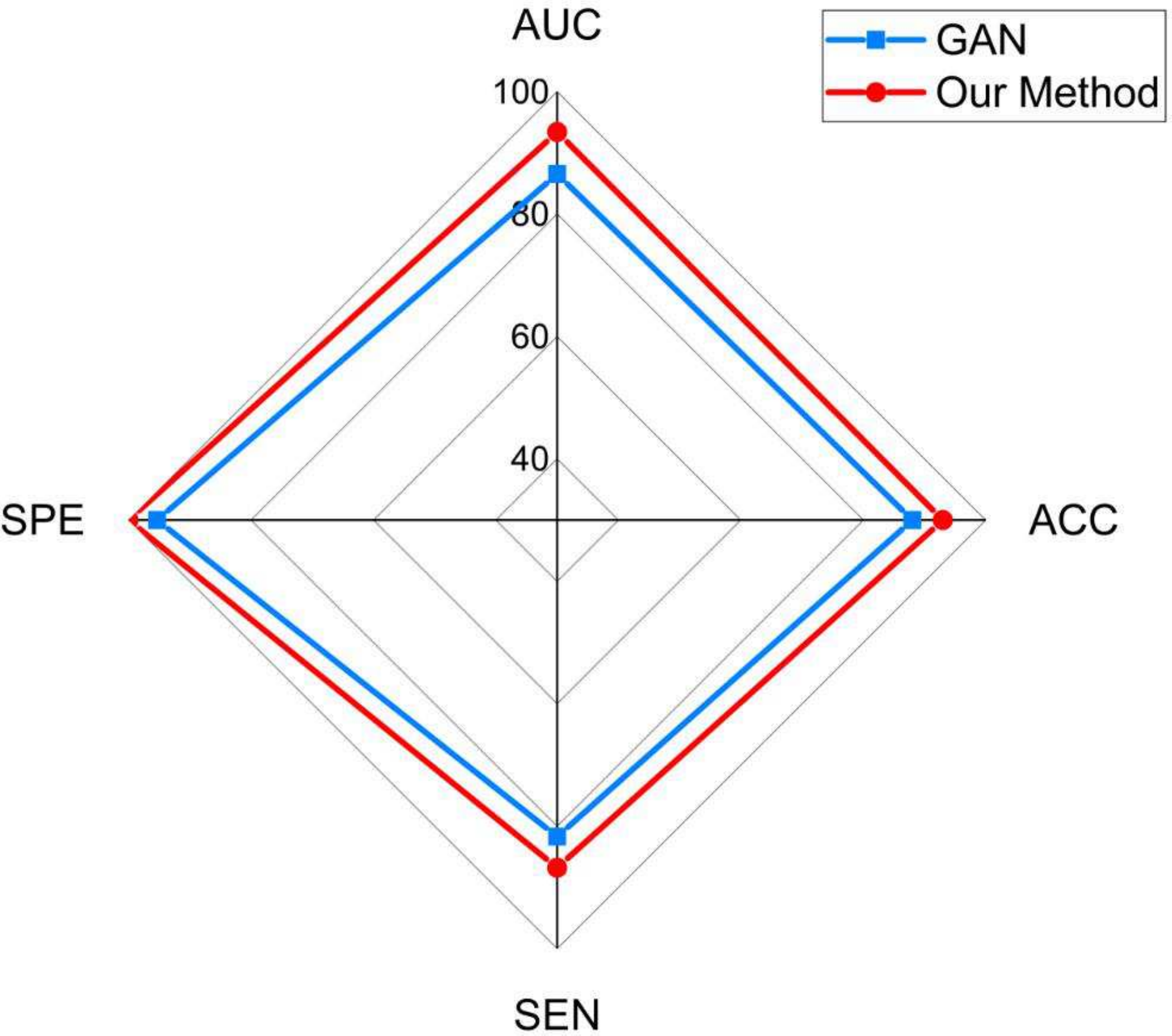}
		\caption{The classification results of synthetic data augmentation for NC vs. SMC.}
		\label{fig_CLASSncSMC}
	\end{figure}
	\begin{figure}
		\centering
		\includegraphics[width=0.9\linewidth]{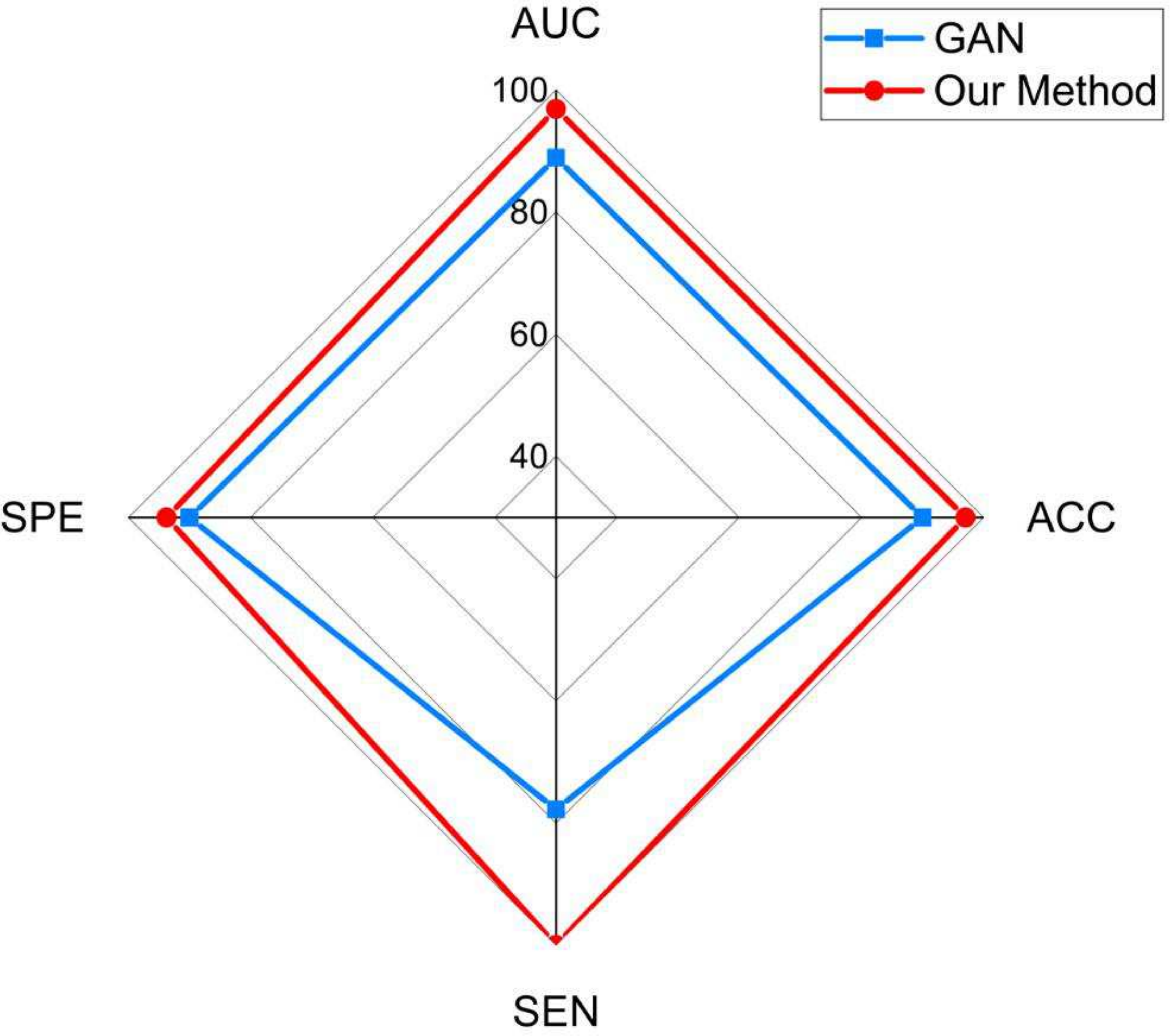}
		\caption{The classification results of synthetic data augmentation for  NC vs. EMCI.}
		\label{fig_CLASSncEMCI}
	\end{figure}
	
	\begin{figure}
		\centering
		\includegraphics[width=0.9\linewidth]{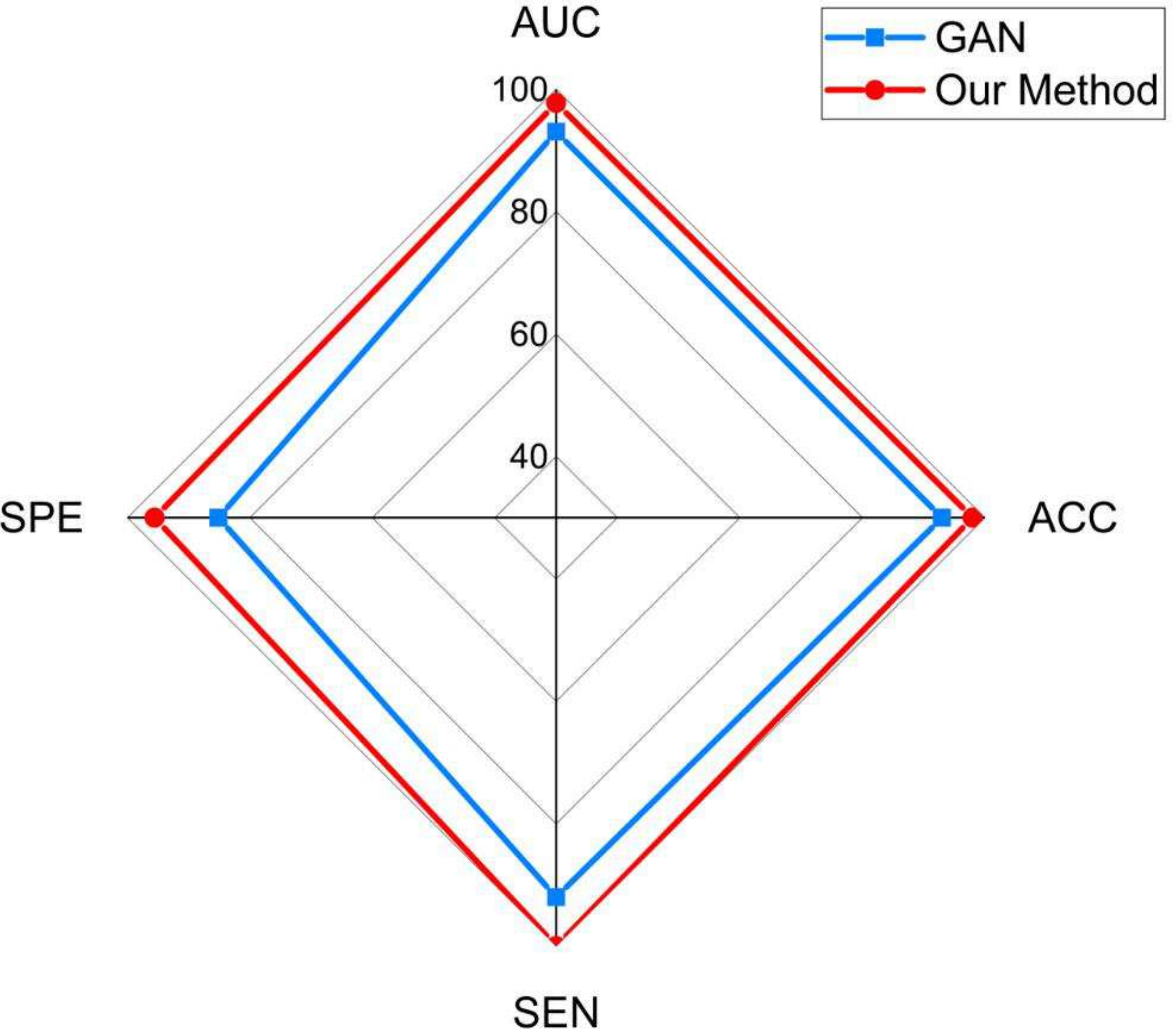}
		\caption{The classification results of synthetic data augmentation for  NC vs. LMCI.}
		\label{fig_CLASSncLMCI}
	\end{figure}

	\begin{figure}
		\centering
		\includegraphics[width=0.9\linewidth]{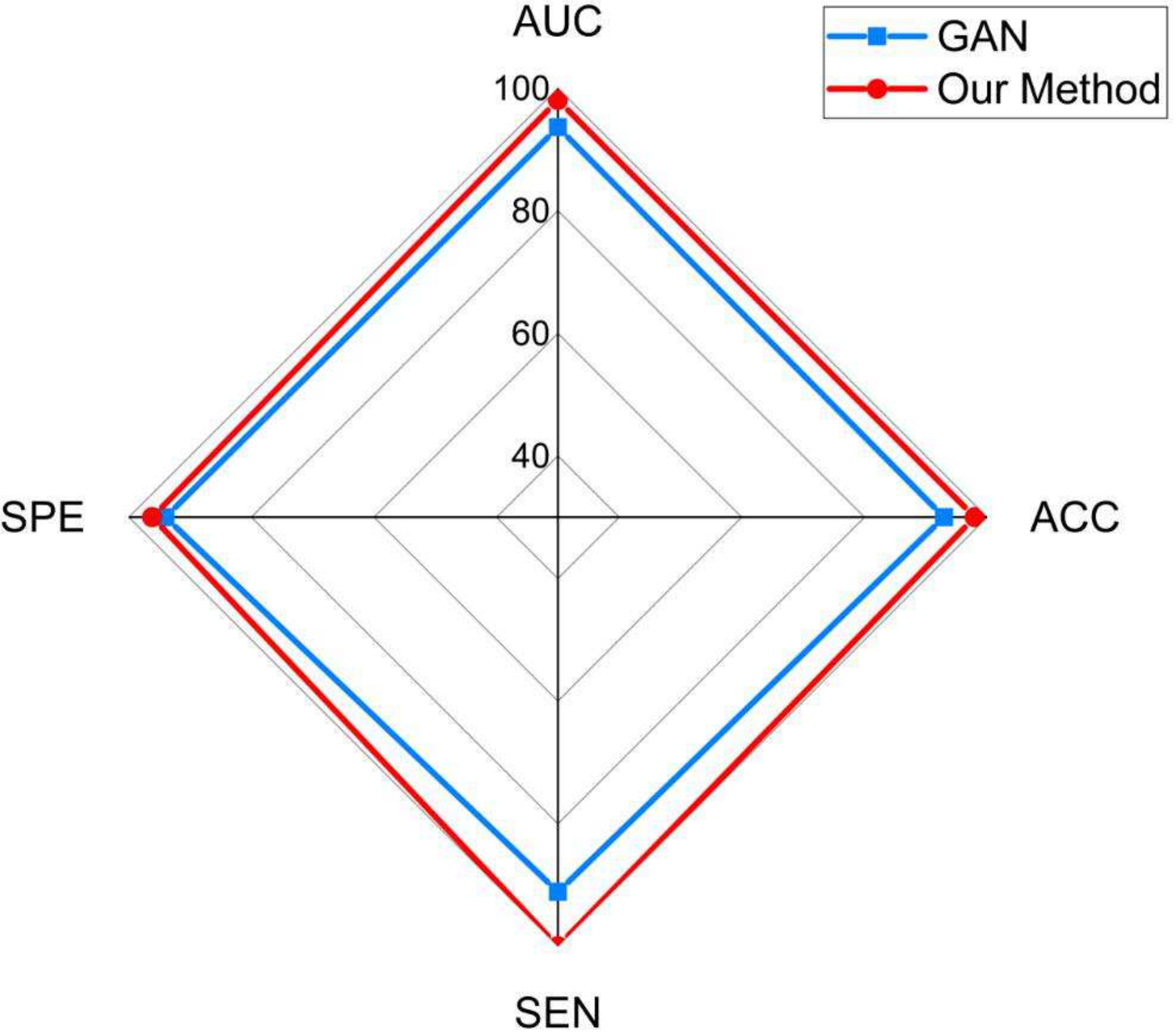}
		\caption{The classification results of synthetic data augmentation for NC vs. AD.}
		\label{fig_CLASSncAD}
	\end{figure}
	
	Furthermore,  the generation diversity with SSIM is evaluated in each iteration on the validation dataset. The convergence curves of the proposed MP-GAN and GAN  are given for 4 evaluation groups: (1)NC vs. SMC, (2)NC vs. EMCI, (3)NC vs. LMCI, and (4)NC vs. AD  respectively. From Fig. \ref{fig_SSIMncSMC} to Fig. \ref{fig_SSIMncAD}, it can be observed that the proposed MP-GAN converges faster than   GAN. Meanwhile,  MP-GAN performs stably in all  4 evaluation groups. On the other hand, the training of  GAN is extremely unstable for NC vs. LMCI, and it can not converge for NC vs. EMCI and NC vs. AD. Again, these results are consistent with   the NCC score in Fig. \ref{fig_NCC}.  The NCC score of GAN is low for NC vs. EMCI in Fig. \ref{fig_NCC}, because GAN can't converge for NC vs. EMCI as shown in Fig. \ref{fig_SSIMncEMCI}.   These results also indicate that the proposed MP-GAN can generate diverse MR images close to the real distribution.

	The objective of the synthetic data augmentation is to demonstrate the learned class-discriminative map have captured all subtle morphological feature for different stages of AD. If the classification performance is improved, this indicates that the learned class-discriminative maps have captured all subtle morphological features for different stages of AD progression. Thus the synthetic MR images produced by the class-discriminative maps can be classified as the corresponding class correctly. By conducting this experiment, the efficacy of the MP-GAN is corroborated.  More specifically, the CNN classifier is trained using synthetic data augmentation. More specifically, the 100 synthesized MR images of each class by  MP-GAN and  GAN are added to the original training set to form two new augmented training sets separately. Then the CNN model is trained on the two new augmented training sets separately for each evaluation group. During the test stage, the same test set of real MR images are used.  From Fig. \ref{fig_CLASSncSMC} to Fig. \ref{fig_CLASSncAD}, it can be seen that adding synthesized samples by the proposed MP-GAN achieves better classification performance in terms of AUC, accuracy, specificity, and sensitivity.   Overall,  the synthetic data samples generated by MP-GAN can add additional variability to the original training set, which in turn leads to better performance. This implies that the synthesized MR images generated by  MP-GAN not only provide meaningful visualizations but also capture the discriminative features for AD analysis.  The proposed MP-GAN can be used as an effective data augmentation method.
	\section{Discussion}\label{sec_Disscussion}
	
	Although the extensive experiments demonstrate the superiority of the proposed MP-GAN. However, MP-GAN has two limitations: (1) The hyperparameters are tuned empirically for the best performance. The optimal value of the hyperparameters depends on network architecture and data. There is no straightforward way to find optimal hyperparameters in advance. (2) The conventional GAN has several common failure modes, such as training instability and mode Collapse. The proposed MP-GAN is designed as a three-player cooperative game instead of the conventional two-player competition game by introducing the auxiliary classifier network based on generator and discriminator. The specific architecture design and the proposed hybrid loss can make the training process more stable. However, mode collapse, which is a common issue of the GAN model, still might happen even MP-GAN has shown stable training performance. In future works, how to solve the mode collapse issue is the direction to further improve the robustness of the proposed MP-GAN.

	\section{Conclusion}\label{sec_Conclusion}
	In this paper,  a novel MP-GAN is proposed to visualize the morphological features indicating the severity of AD in whole-brain MR images.  By introducing a novel multidirectional mapping mechanism into the model, MP-GAN can capture the salient global features efficiently. Thus, by utilizing the class-discriminative map from the generator, the proposed model can clearly delineate the subtle lesions via MR image transformations between the source domain and the target domain. Besides, by integrating the adversarial loss, classification loss, cycle consistency loss, and \emph{L}1 penalty, a single generator in MP-GAN can learn the class-discriminative maps for multiple-classes. Experimental results on the public  ADNI dataset have demonstrated that MP-GAN can visualize multiple lesions affected by the progression of AD accurately.    Furthermore, MP-GAN may visualize some new disease-related regions that have not been investigated yet. This can be studied further to discover potential new AD biomarkers in future work.

	\section*{Acknowledgment}
	This work was supported by the National Natural Science Foundations of China under Grants 62172403 and 61872351, the International Science and Technology Cooperation Projects of Guangdong under Grant 2019A050510030, the Distinguished Young Scholars Fund of Guangdong under Grant 2021B1515020019, the Excellent Young Scholars of Shenzhen under Grant RCYX20200714114641211 and Shenzhen Key Basic Research Projects under Grant JCYJ20200109115641762 and JCYJ20180507182506416, the HKRGC Grant Numbers: GRF 12200317, 12300218, 12300519 and 17201020, and the data support of ADNI.
	\ifCLASSOPTIONcaptionsoff
	\newpage
	\fi
	
	\bibliographystyle{IEEEtran}
	\bibliography{reference}

\begin{thebibliography}{10}
\providecommand{\url}[1]{#1}
\csname url@samestyle\endcsname
\providecommand{\newblock}{\relax}
\providecommand{\bibinfo}[2]{#2}
\providecommand{\BIBentrySTDinterwordspacing}{\spaceskip=0pt\relax}
\providecommand{\BIBentryALTinterwordstretchfactor}{4}
\providecommand{\BIBentryALTinterwordspacing}{\spaceskip=\fontdimen2\font plus
\BIBentryALTinterwordstretchfactor\fontdimen3\font minus
  \fontdimen4\font\relax}
\providecommand{\BIBforeignlanguage}[2]{{%
\expandafter\ifx\csname l@#1\endcsname\relax
\typeout{** WARNING: IEEEtran.bst: No hyphenation pattern has been}%
\typeout{** loaded for the language `#1'. Using the pattern for}%
\typeout{** the default language instead.}%
\else
\language=\csname l@#1\endcsname
\fi
#2}}
\providecommand{\BIBdecl}{\relax}
\BIBdecl

\bibitem{ADpriorRegion2}
X.~Hao, Y.~Bao, Y.~Guo, M.~Yu, D.~Zhang, S.~L. Risacher, A.~J. Saykin, X.~Yao,
  and L.~Shen, ``Multi-modal neuroimaging feature selection with consistent
  metric constraint for diagnosis of alzheimer's disease,'' \emph{Medical Image
  Analysis}, vol.~60, p. 101625, 2020.

\bibitem{ADfactor}
M.~W. Bondi, E.~C. Edmonds, and D.~P. Salmon, ``Alzheimer's disease: Past,
  present, and future,'' \emph{Journal of the International Neuropsychological
  Society : JINS}, vol.~23, no. 9-10, p. 818—831, October 2017.

\bibitem{wang2018bone}
S.~Wang, Y.~Shen, D.~Zeng, and Y.~Hu, ``Bone age assessment using convolutional
  neural networks,'' in \emph{2018 International Conference on Artificial
  Intelligence and Big Data (ICAIBD)}.\hskip 1em plus 0.5em minus 0.4em\relax
  IEEE, 2018, pp. 175--178.

\bibitem{wang2020ensemble}
S.~Wang, X.~Wang, Y.~Shen, B.~He, X.~Zhao, P.~W.-H. Cheung, J.~P.~Y. Cheung,
  K.~D.-K. Luk, and Y.~Hu, ``An ensemble-based densely-connected deep learning
  system for assessment of skeletal maturity,'' \emph{IEEE Transactions on
  Systems, Man, and Cybernetics: Systems}, 2020.

\bibitem{ADnotCure2}
Y.~{Shi}, H.~I. {Suk}, Y.~{Gao}, S.~W. {Lee}, and D.~{Shen}, ``Leveraging
  coupled interaction for multimodal alzheimer’s disease diagnosis,''
  \emph{IEEE Transactions on Neural Networks and Learning Systems}, vol.~31,
  no.~1, pp. 186--200, 2020.

\bibitem{ADx1}
S.~Wang, H.~Wang, Y.~Shen, and X.~Wang, ``Automatic recognition of mild
  cognitive impairment and alzheimers disease using ensemble based 3d densely
  connected convolutional networks,'' in \emph{2018 17th IEEE International
  Conference on Machine Learning and Applications (ICMLA)}, 2018, pp. 517--523.

\bibitem{ADx2}
S.~Wang, Y.~Shen, W.~Chen, T.~Xiao, and J.~Hu, ``Automatic recognition of mild
  cognitive impairment from mri images using expedited convolutional neural
  networks,'' in \emph{International Conference on Artificial Neural Networks},
  2017, pp. 373--380.

\bibitem{lei2020deep}
B.~Lei, M.~Yang, P.~Yang, F.~Zhou, W.~Hou, W.~Zou, X.~Li, T.~Wang, X.~Xiao, and
  S.~Wang, ``Deep and joint learning of longitudinal data for alzheimer's
  disease prediction,'' \emph{Pattern Recognition}, vol. 102, p. 107247, 2020.

\bibitem{hu2019cross}
S.~Hu, J.~Yuan, and S.~Wang, ``Cross-modality synthesis from mri to pet using
  adversarial u-net with different normalization,'' in \emph{2019 International
  Conference on Medical Imaging Physics and Engineering (ICMIPE)}.\hskip 1em
  plus 0.5em minus 0.4em\relax IEEE, 2019, pp. 1--5.

\bibitem{hu2020brain}
S.~Hu, Y.~Shen, S.~Wang, and B.~Lei, ``Brain mr to pet synthesis via
  bidirectional generative adversarial network,'' in \emph{International
  Conference on Medical Image Computing and Computer-Assisted
  Intervention}.\hskip 1em plus 0.5em minus 0.4em\relax Springer, Cham, 2020,
  pp. 698--707.

\bibitem{hu2020medical}
S.~Hu, W.~Yu, Z.~Chen, and S.~Wang, ``Medical image reconstruction using
  generative adversarial network for alzheimer disease assessment with
  class-imbalance problem,'' in \emph{2020 IEEE 6th International Conference on
  Computer and Communications (ICCC)}.\hskip 1em plus 0.5em minus 0.4em\relax
  IEEE, 2020, pp. 1323--1327.

\bibitem{yu2020multi}
S.~Yu, S.~Wang, X.~Xiao, J.~Cao, G.~Yue, D.~Liu, T.~Wang, Y.~Xu, and B.~Lei,
  ``Multi-scale enhanced graph convolutional network for early mild cognitive
  impairment detection,'' in \emph{International Conference on Medical Image
  Computing and Computer-Assisted Intervention}.\hskip 1em plus 0.5em minus
  0.4em\relax Springer, Cham, 2020, pp. 228--237.

\bibitem{lei2022predicting}
B.~Lei, E.~Liang, M.~Yang, P.~Yang, F.~Zhou, E.-L. Tan, Y.~Lei, C.-M. Liu,
  T.~Wang, X.~Xiao \emph{et~al.}, ``Predicting clinical scores for
  alzheimer’s disease based on joint and deep learning,'' \emph{Expert
  Systems with Applications}, vol. 187, p. 115966, 2022.

\bibitem{hu2021bidirectional}
S.~Hu, B.~Lei, S.~Wang, Y.~Wang, Z.~Feng, and Y.~Shen, ``Bidirectional mapping
  generative adversarial networks for brain mr to pet synthesis,'' \emph{IEEE
  Transactions on Medical Imaging}, 2021.

\bibitem{ADnotCure1}
N.~{Mammone}, C.~{Ieracitano}, H.~{Adeli}, A.~{Bramanti}, and F.~C. {Morabito},
  ``Permutation jaccard distance-based hierarchical clustering to estimate eeg
  network density modifications in mci subjects,'' \emph{IEEE Transactions on
  Neural Networks and Learning Systems}, vol.~29, no.~10, pp. 5122--5135, 2018.

\bibitem{ADregressionVisual}
M.~Wang, D.~Zhang, D.~Shen, and M.~Liu, ``Multi-task exclusive relationship
  learning for alzheimer’s disease progression prediction with longitudinal
  data,'' \emph{Medical Image Analysis}, vol.~53, pp. 111 -- 122, 2019.

\bibitem{ADclassificationVisual}
B.~Jie, M.~Liu, and D.~Shen, ``Integration of temporal and spatial properties
  of dynamic connectivity networks for automatic diagnosis of brain disease,''
  \emph{Medical Image Analysis}, vol.~47, pp. 81 -- 94, 2018.

\bibitem{ADregressionVisual99}
G.~S. {Babu} and S.~{Suresh}, ``Sequential projection-based metacognitive
  learning in a radial basis function network for classification problems,''
  \emph{IEEE Transactions on Neural Networks and Learning Systems}, vol.~24,
  no.~2, pp. 194--206, 2013.

\bibitem{ADvisual1}
C.~Lian, M.~Liu, J.~Zhang, and D.~Shen, ``Hierarchical fully convolutional
  network for joint atrophy localization and alzheimer's disease diagnosis
  using structural mri,'' \emph{IEEE Transactions on Pattern Analysis and
  Machine Intelligence}, pp. 50--60, 2018.

\bibitem{gan_goodfellow}
I.~Goodfellow, J.~Pouget-Abadie, M.~Mirza, B.~Xu, D.~Warde-Farley, S.~Ozair,
  A.~Courville, and Y.~Bengio, ``Generative adversarial nets,'' in
  \emph{Advances in Neural Information Processing Systems}, 2014, pp.
  2672--2680.

\bibitem{GanReview}
X.~{Yi}, E.~{Walia}, and P.~{Babyn}, ``{Generative Adversarial Network in
  Medical Imaging: A Review},'' \emph{arXiv e-prints}, Sep. 2018.

\bibitem{mo2009}
L.-F. Mo and S.-Q. Wang, ``A variational approach to nonlinear two-point
  boundary value problems,'' \emph{Nonlinear Analysis: Theory, Methods \&
  Applications}, vol.~71, no.~12, pp. e834--e838, 2009.

\bibitem{wang2009variational}
S.-Q. Wang, ``A variational approach to nonlinear two-point boundary value
  problems,'' \emph{Computers \& Mathematics with Applications}, vol.~58, no.
  11-12, pp. 2452--2455, 2009.

\bibitem{wang2008variational}
S.-Q. Wang and J.-H. He, ``Variational iteration method for a nonlinear
  reaction-diffusion process,'' \emph{International Journal of Chemical Reactor
  Engineering}, vol.~6, no.~1, 2008.

\bibitem{cycle_consistent_loss2}
J.-Y. Zhu, T.~Park, P.~Isola, and A.~A. Efros, ``Unpaired image-to-image
  translation using cycle-consistent adversarial networks,'' in
  \emph{Proceedings of the IEEE International Conference on Computer Vision
  (ICCV)}, Oct 2017.

\bibitem{myGAN}
W.~Yu, B.~Lei, M.~K. Ng, A.~C. Cheung, Y.~Shen, and S.~Wang, ``Tensorizing gan
  with high-order pooling for alzheimer's disease assessment,'' \emph{IEEE
  Transactions on Neural Networks and Learning Systems}, pp. 1--15, 2021.

\bibitem{GANgan}
S.~Wang, X.~Wang, Y.~Hu, Y.~Shen, Z.~Yang, M.~Gan, and B.~Lei, ``Diabetic
  retinopathy diagnosis using multichannel generative adversarial network with
  semisupervision,'' \emph{IEEE Transactions on Automation Science and
  Engineering}, vol.~18, no.~2, pp. 574--585, 2021.

\bibitem{wGAN}
M.~Arjovsky, S.~Chintala, and L.~Bottou, ``{W}asserstein generative adversarial
  networks,'' in \emph{Proceedings of the 34th International Conference on
  Machine Learning}, ser. Proceedings of Machine Learning Research, D.~Precup
  and Y.~W. Teh, Eds., vol.~70.\hskip 1em plus 0.5em minus 0.4em\relax PMLR,
  06--11 Aug 2017, pp. 214--223.

\bibitem{wGAN2}
I.~Gulrajani, F.~Ahmed, M.~Arjovsky, V.~Dumoulin, and A.~C. Courville,
  ``Improved training of wasserstein gans,'' in \emph{NIPS}, 2017, pp.
  5769--5779.

\bibitem{swapTest}
E.~Nigri, N.~Ziviani, F.~Cappabianco, A.~Antunes, and A.~Veloso, ``Explainable
  deep cnns for mri-based diagnosis of alzheimer's disease,'' \emph{arXiv
  e-prints}, 2020.

\bibitem{visualCNN}
M.~D. Zeiler and R.~Fergus, ``Visualizing and understanding convolutional
  networks,'' in \emph{Computer Vision-ECCV 2014}, 2014, pp. 818--833.

\bibitem{classificationVisualResults}
S.~{Korolev}, A.~{Safiullin}, M.~{Belyaev}, and Y.~{Dodonova}, ``Residual and
  plain convolutional neural networks for 3d brain mri classification,'' in
  \emph{2017 IEEE 14th International Symposium on Biomedical Imaging (ISBI
  2017)}, 2017, pp. 835--838.

\bibitem{gradient1}
A.~Mahendran and A.~Vedaldi, ``Salient deconvolutional networks,'' in
  \emph{Computer Vision -- ECCV 2016}, 2016, pp. 120--135.

\bibitem{gradient2}
K.~Simonyan, A.~Vedaldi, and A.~Zisserman, ``Deep inside convolutional
  networks: visualising image classification models and saliency maps.''\hskip
  1em plus 0.5em minus 0.4em\relax ICLR, 2014, pp. 1--8.

\bibitem{gradient69}
J.~Yosinski, J.~Clune, A.~M. Nguyen, T.~J. Fuchs, and H.~Lipson,
  ``Understanding neural networks through deep visualization,'' \emph{arXiv
  e-prints}, vol. abs/1506.06579, 2015.

\bibitem{visualClassifierGradCAMICCV}
R.~R. Selvaraju, M.~Cogswell, A.~Das, R.~Vedantam, D.~Parikh, and D.~Batra,
  ``Grad-cam: Visual explanations from deep networks via gradient-based
  localization,'' in \emph{The IEEE International Conference on Computer Vision
  (ICCV)}, Oct 2017.

\bibitem{gradientAttribution}
M.~Ancona, E.~Ceolini, A.~C. {\"{O}}ztireli, and M.~H. Gross, ``A unified view
  of gradient-based attribution methods for deep neural networks,'' \emph{arXiv
  e-prints}, vol. abs/1711.06104, 2017.

\bibitem{ADvisual7}
M.~Bohle, F.~Eitel, M.~Weygandt, and K.~Ritter, ``Layer-wise relevance
  propagation for explaining deep neural network decisions in mri-based
  alzheimer's disease classification,'' \emph{Frontiers in Aging Neuroscience},
  vol.~11, p. 194, 2019.

\bibitem{GuidedBackprop}
J.~Springenberg, A.~Dosovitskiy, T.~Brox, and M.~Riedmiller, ``Striving for
  simplicity: The all convolutional net,'' in \emph{ICLR}, 2015.

\bibitem{IntegratedGradients}
M.~Sundararajan, A.~Taly, and Q.~Yan, ``Axiomatic attribution for deep
  networks,'' \emph{arXiv e-prints}, vol. abs/1703.01365, 2017.

\bibitem{CAM}
B.~Zhou, A.~Khosla, A.~Lapedriza, A.~Oliva, and A.~Torralba, ``Learning deep
  features for discriminative localization,'' in \emph{The IEEE Conference on
  Computer Vision and Pattern Recognition (CVPR)}, June 2016.

\bibitem{ADCAM}
N.~M. {Khan}, N.~{Abraham}, and M.~{Hon}, ``Transfer learning with intelligent
  training data selection for prediction of alzheimer’s disease,'' \emph{IEEE
  Access}, vol.~7, pp. 72\,726--72\,735, 2019.

\bibitem{ADvisual2}
C.~Lian, M.~Liu, L.~Wang, and D.~Shen, ``End-to-end dementia status prediction
  from brain mri using multi-task weakly-supervised attention network,'' in
  \emph{Medical Image Computing and Computer Assisted Intervention(MICCAI)},
  Jan. 2019, pp. 158--167.

\bibitem{ADfeature}
S.~{Sarraf} and G.~{Tofighi}, ``{Classification of Alzheimer's Disease
  Structural MRI Data by Deep Learning Convolutional Neural Networks},''
  \emph{arXiv e-prints}, p. arXiv:1607.06583, Jul. 2016.

\bibitem{cycle_consistent_loss1}
T.~Kim, M.~Cha, H.~Kim, J.~K. Lee, and J.~Kim, ``Learning to discover
  cross-domain relations with generative adversarial networks,'' \emph{arXiv
  e-prints}, vol. abs/1703.05192, 2017.

\bibitem{ResNet}
K.~He, X.~Zhang, S.~Ren, and J.~Sun, ``Deep residual learning for image
  recognition,'' in \emph{Proceedings of the IEEE Conference on Computer Vision
  and Pattern Recognition (CVPR)}, June 2016.

\bibitem{IN}
D.~Ulyanov, A.~Vedaldi, and V.~S. Lempitsky, ``Instance normalization: The
  missing ingredient for fast stylization,'' \emph{arXiv e-prints}, vol.
  abs/1607.08022, 2016.

\bibitem{denseNet}
G.~Huang, Z.~Liu, L.~van~der Maaten, and K.~Q. Weinberger, ``Densely connected
  convolutional networks,'' in \emph{Proceedings of the IEEE Conference on
  Computer Vision and Pattern Recognition}, 2017.

\bibitem{3D-denseNet}
D.~Gu, ``3d densely connected convolutional network for the recognition of
  human shopping actions,'' in \emph{http://dx.doi.org/10.20381/ruor-21013},
  2017.

\bibitem{BN}
S.~Ioffe and C.~Szegedy, ``Batch normalization: Accelerating deep network
  training by reducing internal covariate shift,'' \emph{arXiv e-prints}, vol.
  abs/1502.03167, 2015.

\bibitem{FSL1}
M.~W. Woolrich, S.~Jbabdi, B.~Patenaude, M.~Chappell, S.~Makni, T.~Behrens,
  C.~Beckmann, M.~Jenkinson, and S.~M. Smith, ``Bayesian analysis of
  neuroimaging data in fsl,'' \emph{NeuroImage}, vol.~45, pp. S173 -- S186,
  2009.

\bibitem{FSL2}
S.~M. Smith, M.~Jenkinson, M.~W. Woolrich, C.~F. Beckmann, T.~E. Behrens,
  H.~Johansen-Berg, P.~R. Bannister, M.~D. Luca], I.~Drobnjak, D.~E. Flitney,
  R.~K. Niazy, J.~Saunders, J.~Vickers, Y.~Zhang, N.~D. Stefano], J.~M. Brady,
  and P.~M. Matthews, ``Advances in functional and structural mr image analysis
  and implementation as fsl,'' \emph{NeuroImage}, vol.~23, pp. S208 -- S219,
  2004.

\bibitem{FSL3}
S.~M. Smith, ``Fast robust automated brain extraction,'' \emph{Human brain
  mapping}, vol.~17, no.~3, p. 143—155, November 2002.

\bibitem{flirt1}
M.~Jenkinson and S.~Smith, ``A global optimisation method for robust affine
  registration of brain images,'' \emph{Medical Image Analysis}, vol.~5, no.~2,
  pp. 143 -- 156, 2001.

\bibitem{flirt2}
M.~Jenkinson, P.~Bannister, M.~Brady, and S.~Smith, ``Improved optimization for
  the robust and accurate linear registration and motion correction of brain
  images,'' \emph{NeuroImage}, vol.~17, no.~2, pp. 825 -- 841, 2002.

\bibitem{VAgan}
C.~F. Baumgartner, L.~M. Koch, K.~C. Tezcan, J.~X. Ang, and E.~Konukoglu,
  ``Visual feature attribution using wasserstein gans,'' \emph{IEEE/CVF
  Conference on Computer Vision and Pattern Recognition}, pp. 8309--8319, 2017.

\bibitem{ccAD}
F.~K. S.~J. Ardekani~BA, Bachman~AH, ``Corpus callosum shape changes in early
  alzheimer's disease: an mri study using the oasis brain database,''
  \emph{Brain Struct Funct}, vol. 219(1), pp. 343--352, 2014.

\bibitem{MMSE}
L.~{Nie}, L.~{Zhang}, L.~{Meng}, X.~{Song}, X.~{Chang}, and X.~{Li}, ``Modeling
  disease progression via multisource multitask learners: A case study with
  alzheimer’s disease,'' \emph{IEEE Transactions on Neural Networks and
  Learning Systems}, vol.~28, no.~7, pp. 1508--1519, 2017.

\bibitem{ADpriorRegion}
Y.~Zhang, Z.~Dong, P.~Phillips, S.~Wang, G.~Ji, J.~Yang, and T.-F. Yuan,
  ``Detection of subjects and brain regions related to alzheimer's disease
  using 3d mri scans based on eigenbrain and machine learning,''
  \emph{Frontiers in Computational Neuroscience}, vol.~9, p.~66, 2015.

\bibitem{MLvisualResults}
J.~M. Rondina, L.~K. Ferreira, F.~L. de~Souza~Duran, R.~Kubo, C.~R. Ono, C.~C.
  Leite, J.~Smid, R.~Nitrini, C.~A. Buchpiguel, and G.~F. Busatto, ``Selecting
  the most relevant brain regions to discriminate alzheimer's disease patients
  from healthy controls using multiple kernel learning: A comparison across
  functional and structural imaging modalities and atlases,'' \emph{NeuroImage:
  Clinical}, vol.~17, pp. 628 -- 641, 2018.

\bibitem{ADpriorRegion1}
C.~{Feng}, A.~{Elazab}, P.~{Yang}, T.~{Wang}, F.~{Zhou}, H.~{Hu}, X.~{Xiao},
  and B.~{Lei}, ``Deep learning framework for alzheimer’s disease diagnosis
  via 3d-cnn and fsbi-lstm,'' \emph{IEEE Access}, vol.~7, pp. 63\,605--63\,618,
  2019.

\bibitem{ADregion1}
H.~Braak and E.~Braak, ``Neuropathological stageing of alzheimer-related
  changes,'' \emph{Acta neuropathologica}, vol.~82, no.~4, p. 239—259, 1991.

\bibitem{ADregion2}
B.~C. Dickerson, A.~Bakkour, D.~H. Salat, E.~Feczko, J.~Pacheco, D.~N. Greve,
  F.~Grodstein, C.~I. Wright, D.~Blacker, H.~D. Rosas, R.~A. Sperling, A.~Atri,
  J.~H. Growdon, B.~T. Hyman, J.~C. Morris, B.~Fischl, and R.~L. Buckner,
  ``{The Cortical Signature of Alzheimer's Disease: Regionally Specific
  Cortical Thinning Relates to Symptom Severity in Very Mild to Mild AD
  Dementia and is Detectable in Asymptomatic Amyloid-Positive Individuals},''
  \emph{Cerebral Cortex}, vol.~19, no.~3, pp. 497--510, 07 2008.

\bibitem{biomarkerNew}
F.~Márquez and M.~A. Yassa, ``Neuroimaging biomarkers for alzheimer's
  disease,'' \emph{Molecular neurodegeneration}, vol.~14, no.~1, p.~21, June
  2019.

\bibitem{BBSI}
N.~C. Fox and P.~A. Freeborough, ``Brain atrophy progression measured from
  registered serial mri: validation and application to alzheimer's disease,''
  \emph{JMRI}, vol.~7, no.~6, pp. 1069--1075, 1997.

\bibitem{SNIPE}
P.~Coupé, S.~F. Eskildsen, J.~V. Manjón, V.~S. Fonov, and D.~L. Collins,
  ``Simultaneous segmentation and grading of anatomical structures for
  patient's classification: Application to alzheimer's disease,''
  \emph{NeuroImage}, vol.~59, no.~4, pp. 3736 -- 3747, 2012.

\bibitem{gradingBiomarker}
T.~{Tong}, Q.~{Gao}, R.~{Guerrero}, C.~{Ledig}, L.~{Chen}, D.~{Rueckert}, and
  A.~D.~N. {Initiative}, ``A novel grading biomarker for the prediction of
  conversion from mild cognitive impairment to alzheimer's disease,''
  \emph{IEEE Transactions on Biomedical Engineering}, vol.~64, no.~1, pp.
  155--165, 2017.

\bibitem{PSNR}
A.~Horé and D.~Ziou, ``Image quality metrics: Psnr vs. ssim,'' in \emph{2010
  20th International Conference on Pattern Recognition}, 2010, pp. 2366--2369.

\bibitem{SSIM}
Z.~{Wang}, E.~P. {Simoncelli}, and A.~C. {Bovik}, ``Multiscale structural
  similarity for image quality assessment,'' in \emph{The Thrity-Seventh
  Asilomar Conference on Signals, Systems Computers}, vol.~2, 2003, pp.
  1398--1402.

\end{thebibliography}
	
\end{document}